\documentclass[useAMS,usenatbib]{mn2e}
\bibpunct[, ]{(}{)}{;}{a}{}{,}

\newcommand\ksmpc{{~km~s$^{-1}$~Mpc$^{-1}$}}
\newcommand\hda{H$\delta_\mathrm{A}$\xspace}
\newcommand\dfk{{$\mathrm{D_n4000}$}\xspace}
\newcommand\fa{{$f_\mathrm{young}$}\xspace}
\newcommand\py{{$P_\mathrm{young}$}\xspace}

\newcommand\ion[2]{#1$\,${\scshape{#2}}}
\newcommand\oiii{[\ion{O}{iii}]$\lambda$5007\xspace}
\newcommand\oii{[\ion{O}{ii}]\xspace}
\newcommand\nii{[\ion{N}{ii}]$\lambda$6584\xspace}
\newcommand\ha{H$\alpha$\xspace}
\newcommand\hb{H$\beta$\xspace}
\newcommand\logm{$\log\,(M/M_\odot)$\xspace}
\newcommand\sn{$S/N$\xspace}

\usepackage{xspace}
\usepackage{amssymb}
\usepackage{graphicx}
\usepackage{booktabs}
\usepackage[colorlinks=True, citecolor=blue]{hyperref}
\usepackage{hyperref}
\usepackage{txfonts}

\defcitealias{brinchmann2004}{B04}
\defcitealias{stasinska2006}{S06}
\defcitealias{kauffmann2003}{K03}
\defcitealias{kewley2001}{K01}

\begin{document}

\title
[The photometric properties of quenched galaxies] 
{Towards a physical picture of star-formation quenching: the photometric properties of recently-quenched galaxies in the Sloan Digital Sky Survey} 

\author
[Mendel et al.]
{J. Trevor Mendel$^{1,2,}$\thanks{jtmendel@mpe.mpg.de}, Luc Simard$^3$, Sara L. Ellison$^1$, David R. Patton$^4$\\ 
$^1$Department of Physics and Astronomy, University of Victoria, Victoria, British Columbia, V8P 1A1, Canada\\
$^2$Max-Planck-Institut f\"{u}r Extraterrestrische Physik, Giessenbachstra\ss e, D-85748 Garching, Germany\\
$^3$National Research Council of Canada, Herzberg Institute of Astrophysics, 5071 West Saanich Road, Victoria, British Columbia, V9E 2E7, Canada\\
$^4$Department of Physics \& Astronomy, Trent University, 1600 West Bank Drive, Peterborough, Ontario, K9J 7B8, Canada}

\maketitle

\begin{abstract}

We select a sample of young passive galaxies from the Sloan Digital Sky Survey Data Release 7 in order to study the processes that quench star formation in the local universe.  Quenched galaxies are identified based on the contribution of A-type stars to their observed (central) spectra and relative lack of ongoing star formation; we find that such systems account for roughly 2.5 per cent of all galaxies with \logm $\geq 9.5$, and have a space density of $\sim$$2.2\times10^{-4}~\mathrm{Mpc}^{-3}$.  We show that quenched galaxies span a range of morphologies, but that visual classifications suggest they are predominantly early-type systems.  Their visual early-type classification is supported by quantitative structural measurements (S\'ersic indices) that show a notable lack of disk-dominated galaxies, suggesting that any morphological transformation associated with galaxies' transition from star-forming to passive---e.g. the formation of a stellar bulge---occurs contemporaneously with the decline of their star-formation activity.  We show that there is no clear excess of optical AGN in quenched galaxies, suggesting that: i) AGN feedback is not associated with the majority of quenched systems or ii) that the observability of quenched galaxies is such that the quenching phase in general outlives any associated nuclear activity.  Comparison with classical post-starburst galaxies shows that both populations show similar signatures of bulge growth, and we suggest that the defining characteristic of post-starburst galaxies is the \emph{efficiency} of their bulge growth rather than a particular formation mechanism.

\end{abstract}


\section{Introduction}
\label{intro}

Large surveys such as the Sloan Digital Sky Survey \citep[SDSS;][]{york2000} and 2dF Galaxy Redshift Survey \citep{colless2001} have placed tight constraints on properties of the local galaxy population and, increasingly, spectroscopic surveys out to $z\sim1$ (e.g. DEEP2 [\citealp{davis2003}] and zCOSMOS [\citealp{lilly2007}]) inform our phenomenological picture of how galaxies have evolved over the past 8 billion years.  For example, we know that the local galaxy population is bimodal in colour: passive galaxies form a tight `red sequence' in the colour--stellar mass plane while star-forming galaxies populate a more dispersed `blue cloud' \citep[e.g.][]{strateva2001,baldry2004}, and these sequences are in place already by at least $z=1$ \citep[e.g.][]{cirasuolo2007,cooper2008a,williams2009,brammer2009,muzzin2012}.  The number density of red-sequence galaxies is observed to evolve significantly from $z\sim1$ to the present \citep{bell2004,faber2007}, suggesting that star-formation quenching plays an important role in galaxy evolution over this epoch.

However, it does not appear that simple evolutionary scenarios, e.g. gas exhaustion, can explain the transition of galaxies from star-forming to quiescent in the absence of additional physical processes.  There is a correlation between star formation and galaxy structure such that passive galaxies are typically spheroid dominated, while star-forming galaxies are generally disky \citep[e.g.][but see also \citealp{masters2010}]{bamford2009,skibba2009}, and both star formation and morphology are connected to galaxies' environment \citep[e.g.][and references therein]{lewis2002,goto2003,kauffmann2004,balogh2004}.  The buildup of passive galaxies since $z\sim$~1 appears to be mirrored by a change in galaxies' light profiles \citep[e.g.][]{mcgee2008,mcgee2011,bell2012}, and therefore the evolution of both star-formation rate and morphology must occur in such a way as to jointly preserve the relationships between galaxy structure, star formation and environment from high redshift to the present.

Numerous physical mechanisms have been proposed to explain galaxies' transition from star-forming disks to passive spheroids with time, both in terms of their internal properties and their surrounding environments.  Mergers are thought to play an important role in both assembling stellar mass and triggering efficient star formation in groups and the field, where low encounter velocities are favourable to strong tidal interactions.   Major mergers (between equal-mass galaxies) can rapidly transform disk-dominated galaxies into spheroids \citep[but see also \citealp{hopkins2009}]{toomre1972,negroponte1983,mcintosh2008}, while minor mergers promote bulge growth by exciting instabilities in otherwise stable disks.  There is evidence that tidal interactions can drive radial gas flows in disks, leading to centrally-concentrated starbursts \citep{mihos1994,ellison2010,scudder2012a}.  Such inflows can also fuel accretion onto a central supermassive black hole; subsequent energetic feedback may be sufficient to prevent additional gas cooling \citep[e.g.][]{kauffmann2000,croton2006,hopkins2008,van-de-voort2011}, efficiently smothering star formation.  Environmental processes tied either to galaxies' interaction with the intra-cluster medium \citep[i.e. ram-pressure or viscous stripping;][]{gunn1972,nulsen1982} or the properties of their host dark-matter haloes \citep[so-called ``halo quenching'';][]{keres2005,dekel2006} may also play an important part in regulating star-formation activity.  Increasingly, a combination of external mechanisms and internal, secular processes are also invoked in order to reproduce the observed properties of the galaxy population.  Disk instabilities can lead to the formation of a bar \citep[e.g.][and references therein]{sellwood1993} which can, in turn, drive radial gas flows and the growth of a secular (pseudo) bulge. Observations in the local universe suggest that this secular bulge growth is a nearly ubiquitous feature in disk-dominated systems \citep{kormendy2010}. 

Confronting our theories of galaxy evolution with observational data is a critical step in developing a refined, physical understanding of the galaxy population as we see it today.  Wholesale comparison of the observed and theoretical galaxy population has been significantly aided by the development of so-called semi-analytic models \citep[e.g.][]{kauffmann1993,cole1994,somerville1999}, which `paint' galaxies onto high resolution $N$-body simulations using analytic prescriptions to capture the behaviour of relevant physical processes.  Such models enjoy success in reproducing the bulk properties of galaxies at $z\sim0$, such as the luminosity and stellar mass functions \citep[e.g.][]{bower2006,de-lucia2007a} and the colour--magnitude relation \citep[e.g.][]{weinmann2006a,henriques2009}.  However, despite these successes, several details of the galaxy population are poorly reproduced.  Galaxies assemble too quickly, leading to a significant over-prediction of the stellar mass function at $z>1$ \citep[e.g.][]{somerville2008,guo2011} and an offset between the ages of simulated and observed galaxies, particularly at low mass \citep[e.g.][]{fontanot2009,weinmann2012}. Environmental processes are too efficient, resulting in an overproduction of passive low-mass galaxies in high-density environments by up to 50 per cent \citep[e.g.][]{weinmann2006a,kimm2009}.  More generally, models seem to under-predict star-formation rates by a factor of $\sim$4 at $z < 1$ \citep[e.g.][]{daddi2007}, suggesting our understanding of mechanisms that regulate star formation is incomplete.

One way to further constrain formation of the passive population is to study directly the properties of galaxies in transition between star-forming and quiescent; however, such galaxies are expected to be rare in the local Universe.  Nevertheless, samples of so-called ``K+A'', ``E+A'' or ``post-starburst'' galaxies\footnote{For simplicity we will refer to K+A, E+A and post-starburst galaxies collectively as post-starburst galaxies, though we note that this is not an observational definition, relying instead on the interpretation of such galaxies' spectra as arising from a starburst.} can be identified based on the presence of strong Balmer-line absorption---characteristic of A-type stars---and a lack of optical emission lines, and have recent star-formation histories characterised by intense star-formation followed by a precipitous decline.  While such galaxies were first studied in the context of clusters \citep[e.g.][]{couch1987}, they have since been shown to inhabit a broad range of environments \citep{zabludoff1996,blake2004,hogg2006,nolan2007}, supporting the idea that their formation is not necessarily governed by cluster-specific processes.  Indeed, photometric observations have shown that anywhere from 30 to 80 percent of post-starburst galaxies are morphologically disturbed \citep{yang2004a,goto2005,pracy2009}, highlighting mergers as one of the likely drivers of post-starburst galaxy formation.  However, the high burst-mass fractions inferred for some post-starburst galaxies, anywhere from 20--60 per cent \citep[e.g.][]{kaviraj2007}, suggest that they be biased towards relatively extreme formation scenarios, and may not reflect the evolution of ``average'' galaxies.

In this paper we exploit a wealth of SDSS data to identify a large sample of galaxies in transition between star forming and passive, hereafter referred to as `quenched', facilitating a direct study of their properties.  This work builds upon the substantial body of literature characterising the \emph{integrated} properties of the passive galaxy population \citep[e.g. colour or star-formation rate;][]{weinmann2006,van-den-bosch2008,wetzel2012} by focusing on a subsample of passive galaxies that have only recently stopped forming stars.  In studying the frequency, physical properties and environments of such systems our goal is to constrain the processes that lead to their formation and, more generally, the physical mechanisms that contribute to the buildup of the red (passive) sequence over time.  In this paper we identify quenched galaxies based on presence of relatively young stellar populations coupled with low instantaneous star-formation rates (SFRs), and show that quenched systems bear a close resemblance to the population of galaxies already on the red sequence.  In a forthcoming paper (Mendel et al. in prep.) we will discuss more directly the relationship between quenched galaxies and their host environments.  In Section \ref{data} we outline the characteristics of our SDSS sample, while in Section \ref{quenched_select} we detail our selection of quenched galaxies.  Readers less interested in the technical aspects of our sample selection may wish to begin with Section \ref{results}, where we focus on comparing the properties of quenched galaxies with other populations in the SDSS.

Throughout this paper we adopt a flat cosmology with $\Omega_{\Lambda} = 0.7$, $\Omega_{\mathrm{M}} = 0.3$ and $H_0 = 70$\ksmpc.  All magnitudes are given in the AB system, and we adopt a notation such that equivalent widths in emission are negative.

\section{Data}
\label{data}

This work is based on data from the SDSS, supplemented by several value-added galaxy catalogues.  We adopt photometric measurements from the catalogues of \citet{simard2011}, and galaxy stellar masses are estimated from fits to galaxies' $ugri$ spectral energy distributions (SEDs) as described by \citet{mendel2012a}.  We adopt measurements of specific star-formation rate (SSFR $\equiv$ SFR/$M_\star$) from the MPA/JHU data catalogues \footnote{accessed at \url{http://www.mpa-garching.mpg.de/SDSS/DR7/}}.  In order to provide a more robust classification of galaxies' morphology we supplement the purely quantitative morphological parameters estimated by \citet{simard2011} with visual classifications from the Galaxy Zoo first data release\footnote{accessed at \url{http://data.galaxyzoo.org}.} \citep{lintott2011}.  Below we briefly summarise the relevant characteristics of these catalogues.

\subsection{Galaxy sample}
\label{sdss_sample}

The sample used here is selected from the SDSS Data Release 7 \citep[DR7;][]{abazajian2009} to include all objects with extinction-corrected $r$-band Petrosian magnitudes, $m_r$, satisfying $14 < m_r \leq 17.77$, and classified as galaxies both photometrically and spectroscopically by the SDSS data reduction pipelines (i.e. {\tt photoObj.type = 3} and {\tt specObj.specClass = 2}).  We further require that all spectra have a redshift confidence {\tt zConf $\geq 0.9$}.  Our identification of quenched galaxies requires an estimate of the current star-formation rate, and leads us to consider only those galaxies with $z \leq 0.2$, ensuring that \ha stays blue-ward of significant night-sky emission at $\lambda > 8000$\AA.  We also adopt a lower redshift limit of $z = 0.01$ to ensure that distance measurements are cosmological.

\subsection{Photometric properties}
\label{gim2d}

\citet{simard2011} performed bulge-disk decompositions for galaxies in the SDSS DR7, resulting in a catalogue of structural properties for $\sim$1.12 million objects with extinction-corrected $r$-band Petrosian magnitudes $14 < m_r \le 18$.  A full description of these decompositions can be found in \citet{simard2011}; below we provide an outline of their methodology to highlight relevant details.

\citet{simard2011} reprocessed the corrected SDSS images (output by the {\sc frames} pipeline), using {\sc sextractor} \citep{bertin1996} to identify objects and generate their associated segmentation images.  Local sky levels were then re-determined on an object-by-object basis using a minimum of 20,000 sky pixels, excluding any pixels within $4^{\prime\prime}$ of any primary or neighbouring object mask.  Structural decompositions were performed using the {\sc gim2d} software package \citep{simard2002} and three different PSF-convolved axisymmetric models: (i) a single-component S\'ersic profile, (ii) a two-component de Vaucouleurs bulge plus exponential disk and (iii) a two-component S\'ersic bulge plus exponential disk.  In all cases fits were performed simultaneously in the $g$ and $r$ bands by forcing bulge (half-light size, ellipticity and position angle) and disk (scale length, inclination and position angle) structural properties to be the same in both bandpasses; remaining parameters---e.g. total flux, bulge-to-total flux ratio (B/T) and model centroid relative to the SDSS object position\footnote{\citeauthor{simard2011} found that small errors in the image registration between the $g$ and $r$ bands resulted in a significant number of galaxies with ``positive--negative'' residuals in their cores.}---were fit independently in each bandpass.  

In the present work we are primarily interested in galaxies' global properties, and therefore adopt photometric and structural parameters from single-component S\'ersic profile fits.  All rest-frame quantities are computed at $z=0$ by using {\sc k-correct v4\verb|_|2} \citep{blanton2007a} and have been corrected for foreground (Galactic) extinction using the reddening maps of \citet{schlegel1998}.

\subsection{Visual morphologies}
\label{zoo_morph}

The quantitative morphologies described in Section \ref{gim2d} allow for analyses of galaxy structural properties, however it is also helpful to relate these quantitative measurements to galaxies' apparent visual morphology.  In general, visual classifications of galaxy morphology are carried out by a small number of professionally-trained astronomers.  While this approach ultimately produces robust estimates of galaxy morphology, it is hamstrung by relying upon a small pool of identifiers to produce consistent morphological classes across an entire catalogue. 

In order to circumvent this limitation the Galaxy Zoo project \citep{lintott2008,lintott2011} has taken a distributed approach to morphological classification, leveraging public interest to visually classify $\sim$900,000 galaxies from the SDSS.  Whereas classical morphological typing hinges on a small number of accurate visual types, Galaxy Zoo relies on a significant number of less-certain classifications per galaxy to provide probabilistic estimates of galaxy morphology.  These data have already been used to explore many aspects of the relationship between galaxies and their host environments \citep[e.g.][]{skibba2009,bamford2009,darg2010a,darg2010,masters2010}.  Galaxy Zoo asked users to place galaxies into one of four categories: elliptical, spiral, merger and ``don't know''.  We use the de-biased elliptical (E) and combined spiral (CS)\footnote{Galaxy Zoo asks participants to classify spiral galaxies into sub-classes of clockwise, anti-clockwise and edge-on, all three of which are included in the combined spiral classification.} probabilities, which account for the preferential classification of galaxies as early type with increasing redshift (see \citealp{bamford2009} and \citealp{lintott2011} for details of the de-biasing procedure).

\subsection{Stellar masses}

\citet{mendel2012a} provide stellar mass estimates for galaxies in the \citet{simard2011} catalogue based on fits to their $ugri$ SEDs.  These fits are conducted using a large grid of synthetic photometry which encompasses a range of plausible age, metallicity, star-formation history and dust content.  Models are generated using the flexible stellar population synthesis (FSPS) code of \citet{conroy2009}, adopting a \citet{chabrier2003} stellar initial mass function (IMF).  Estimates of the stellar mass are taken as the median of the (marginalized) posterior probability density function, with uncertainties quoted as the 16th and 84th percentiles of the distribution.  Statistical uncertainties on the derived stellar masses are of order 0.15 dex, and \citeauthor{mendel2012a} show that reasonable variations in the `default' FSPS parameters lead to systematic offsets of less than $\pm0.1$ dex in stellar mass for most galaxies.  We adopt here masses from the \citet{mendel2012a} catalogue based on the S\'ersic profile fits of \citet{simard2011}.

\subsection{Star-formation rates}
\label{sfr}

\citet[hereafter B04]{brinchmann2004} estimate star-formation rates for SDSS galaxies using a combination of detailed spectral modelling and empirically-determined calibrations to account for contamination from non-stellar ionising sources, e.g. active galactic nuclei (AGN).  The adopted approach depends on the relative strength and signal-to-noise (\sn) of the \hb, \oiii, \ha and \nii emission lines, which we describe in more detail below.

Galaxies with well-detected emission lines ($S/N > 3$) are first classified as either star forming, composite or AGN using the \oiii/\hb versus \nii/\ha line diagnostic of \citet[hereafter BPT]{baldwin1981} and the demarcations of \citet{kauffmann2003} and \citet{kewley2001}; roughly 40 per cent of galaxies are classifiable in this way.  The remaining 60 per cent are therefore classified based on reduced sets of emission lines.  Galaxies with $S/N > 3$ in both \ha and \nii, and \nii/\ha $>$ 0.6 (8 percent of the sample) are classified as low-\sn AGN.  Remaining galaxies with $S/N > 2$ in \ha (20 per cent) are classified as low-\sn star forming, while galaxies failing all of the above criteria are unclassifiable on the basis of their emission lines.

SFRs for star-forming galaxies are determined from detailed fitting of their emission lines using the models of \citet{charlot2001}.  \citetalias{brinchmann2004} compute the full probability density function of a given parameter (e.g. star-formation rate, metallicity, etc.) using a Bayesian approach to determine the likelihood of each point in a grid of models given the observed data.  Although this fitting approach incorporates information from multiple emission lines, in practice star-formation rates are driven primarily by \ha luminosity.

All other galaxies---i.e. those classified as AGN, composite or unclassifiable---have SFRs estimated from an empirical relationship between 4000\AA~break strength \citep[\dfk;][]{balogh1999} and SSFR based on the star-forming galaxy sample.  The advantage of this approach is that it allows the determination of star-formation rates even for galaxies whose emission-line spectra are heavily contaminated by AGN.  However, it is important to note that, since \dfk is primarily correlated with luminosity-weighted age rather than SSFR, adopting the empirical relationship between SSFR and \dfk of star-forming galaxies automatically assumes that their recent star-formation histories are representative of all galaxies with a given 4000\AA~break strength.  This assumption is particularly misleading for galaxies with discontinuous star-formation histories, as is the case for so-called `post-starburst' spectral types \citep[e.g.][]{zabludoff1996,goto2003}.  We will revisit this issue in Section \ref{quenched_select}.

For the majority of this work we limit ourselves to using SSFRs determined within the SDSS fibre aperture in order to be consistent with our own measurements of galaxies' stellar populations (Section \ref{quenched_select}); however, \citetalias{brinchmann2004} also provide estimates of galaxies' total SFR accounting for star formation occurring outside of the 3$^{\prime\prime}$ SDSS fibre aperture.  In general, there is good agreement between \citetalias{brinchmann2004}'s aperture-corrected star-formation rates and estimates of the total galaxy star-formation rate obtained from $UV$+optical SED fitting \citep{salim2007}.

\subsection{Sampling and volume corrections}
\label{sampling}

In order to discuss absolute quantities we have to correct for the visibility of galaxies in our sample, which can be encapsulated by two factors: the effect of our photometric selection (Section \ref{sdss_sample}) on the sample volume and the impact of our \sn requirements on the spectral sampling rate.  

First, galaxies will be observable within some volume $V_\mathrm{max}$ which depends on their SED shape and overall luminosity.  Along with their stellar mass measurements, \citet{mendel2012a} provide estimates of the minimum and maximum observable redshift ($z_\mathrm{min,SED}$ and $z_\mathrm{max,SED}$) for each galaxy on the basis of their best-fit SED\footnote{Note that the minimum observable redshift arises because the \citet{simard2011} photometric catalogue excludes galaxies with $m_r \leq 14$.}.  We use these estimates along with our adopted redshift limits ($0.01 \leq z \leq 0.2$) to compute the final volume limit, which is constrained by

\begin{equation}
z_\mathrm{min} = \mathrm{max}(z_\mathrm{min,SED}, 0.01),
\end{equation}

\begin{equation}
z_\mathrm{max} = \mathrm{min}(z_\mathrm{max, SED}, 0.2) 
\end{equation}

\noindent and the survey area.  In what follows we correct for these volume effects by weighting galaxies according to their observable volume as $V_\mathrm{max}^{-1}$ \citep[e.g.][]{schmidt1968}.

Second, in order to characterise galaxies' stellar populations we require that their spectra have a \sn of at least 10 in the wavelength range of interest, 3800--5400\AA~(see Section \ref{reliability}).  For each galaxy we estimate \sn based on unmasked pixels used in our spectral decomposition and the SDSS error spectra.  The detectability of a given galaxy in the SDSS fibre depends on its apparent magnitude, concentration\footnote{$C \equiv R90/R50$, where $R90$ and $R50$ are defined as the radii enclosing 90 and 50 per cent of the $r$-band Petrosian flux} and redshift.  In order to correct for the spectral sampling rate we compute the fraction of galaxies within a window of $\Delta z = 0.01$, $\Delta C = 0.2$ and $\Delta m_g = 0.2$ that meet our \sn requirements, $f_{S/N}$, and weight galaxies as $w_\mathrm{SSR} \equiv f_{S/N}^{-1}$.  The majority of galaxies (94 per cent) have $w_\mathrm{SSR} \leq 2$.  In the following, all data are weighted by $w_\mathrm{SSR}V_\mathrm{max}^{-1}$ to account for sampling effects unless otherwise stated.

\section{Selecting quenched galaxies}
\label{quenched_select}

Our goal in this work is to study the most recent additions to the passive galaxy population; in simple terms, this goal can be met by identifying galaxies that are young but host little or no ongoing star formation.  These criteria can be thought of as analogous to the selection of post-starburst galaxies, which is typically based on the presence of strong Balmer-line absorption and relative lack of optical emission lines \citep[e.g.][]{couch1987,zabludoff1996,goto2003,blake2004,wong2012}, generalised to identify a more inclusive sample of quenched systems.  We will discuss a more direct comparison with post-starburst galaxies in Section \ref{psb_comp}. 

We first identify galaxies that host little or no ongoing star formation.  As discussed in Section \ref{sfr}, the \citetalias{brinchmann2004} SFRs are only computed directly from the \ha emission-line flux for galaxies classified as star forming on the BPT diagram, or with \ha $S/N > 2$ once the majority of AGN galaxies are excluded; remaining galaxies have their SFR estimated from \dfk.  In terms of classifying quenched galaxies this is problematic: while \ha emission is sensitive to the lifetime of \ion{H}{ii} regions, $\sim$$10^7$~yrs, \dfk traces the luminosity-weighted age of the population, which is sensitive to the \emph{integrated} mass of young stars on timescales over which the stellar mass-to-light ratio evolves significantly, i.e. hundreds of Myrs.  

Despite the above complication, pursuing a selection based on SSFR has two benefits.  First, there is a tight relationship between SFR and mass for star-forming galaxies out to at least $z=1$ \citep[e.g.][]{noeske2007}, which suggests that adopting a division between star-forming and passive galaxies based on their instantaneous SFR (or in this case, SSFR) offers a robust selection at any redshift.  Second, while the canonical selection of post-starburst galaxies is based on the absence of particular emission lines, typically \oii or \ha, \citet{yan2006} have shown that the genesis of emission lines, particularly \oii, in many post-starburst galaxies is \emph{not} residual star formation but low-ionisation nuclear emission-line region (LINER) activity.  Selecting quenched galaxies based on the absence of \oii emission may therefore exclude a significant fraction of `true' quenched systems.  While a more general selection of quenched galaxies can be obtained based on \ha equivalent width \citep{yan2006}, it is important to note that samples selected on the absence of \ha still exclude strong-line AGN and almost exclusively sample the population of LINER-like galaxies \citep{cid-fernandes2010,cid-fernandes2011}.  In addition, it has been argued that a significant fraction of observed LINER-like activity may be due to photoionisation by hot evolved stars (e.g. white dwarfs and post-asymptotic giant branch [AGB] stars), which could account for the extended LINER-like emission observed in many early-type galaxies \citep[e.g.][]{stasinska2008,sarzi2010,yan2012}.  In this case we might actually expect an \emph{excess} of LINER systems within $\sim$1~Gyr of having their star formation quenched, as stars in the range 3--6 M$_\odot$ contribute significantly to the post-AGB population \citep[e.g.][]{taniguchi2000,stasinska2008,cid-fernandes2011}.  

We therefore adopt a definition of passive galaxies based on both the \citetalias{brinchmann2004} SSFR and observed EW(\ha) in an effort to be as inclusive as possible of potential quenched galaxies.  We begin by adopting a relationship between fibre SSFR and fibre stellar mass for star-forming galaxies in the \citetalias{brinchmann2004} sample, which is well described by $\mathrm{SSFR} \propto \mathrm{M_\star}^{\beta}$ with $\beta = -0.16$.  We identify 329,341 galaxies with SSFRs a factor of 10 below this fiducial sequence ($\sim$3$\sigma$ offset from the SSFR--mass relation), corresponding to a cut in \dfk of $\sim$1.6.  We then supplement this passive sample by including an additional 14,874 galaxies with EW(\ha) $\geq -3$\AA~not identified as star forming or low-\sn star forming by \citetalias{brinchmann2004}\footnote{\citet{salim2007} have shown that there is good agreement between their $UV$-derived star-formation rates and those of \citetalias{brinchmann2004} for both star-forming and low-\sn star-forming galaxies with \dfk $>1.7$, while for \dfk $< 1.7$ \citetalias{brinchmann2004} may overestimate the SFRs of low-\sn star-forming galaxies.  As our adopted division in SSFR can be effectively translated into a cut in \dfk $>1.6$, any remaining low-\sn galaxies are likely to be genuinely star forming.}, corresponding to the EW(\ha) limit described by \citet{cid-fernandes2011} for selecting `retired' galaxies.

\subsection{Quantifying ``young''}
\label{py}

Having outlined the basis for our selection of passive galaxies above, we now turn to our classification of galaxies as ``young''.  As discussed previously, our method for classifying galaxy stellar populations can be thought of as an extension of criteria used to select post-starburst galaxies, which typically use cuts on Balmer-line equivalent width (EW), e.g. $\mathrm{EW(H\delta) \geq 5}$\AA, to identify those galaxies hosting significant A-star populations.  However, adopting such strict cuts to identify truncated star-formation histories primarily selects a subset of quenched galaxies with the highest fraction of young stars and most rapidly declining star-formation rates \citep[e.g.][]{snyder2011}, and it is unclear how representative these are of the \emph{majority} of galaxies in transition between star forming and passive.  More recently, various spectral decomposition techniques have been used to identify galaxies hosting even relatively minor populations of young stars, resulting in a more inclusive sampling of the transition population \citep[e.g.][]{ellingson2001,quintero2004,yan2006,nolan2007,wild2009,yan2009}.  It is this latter approach that we adopt here.

\begin{figure}
\centering
\includegraphics[scale=0.73]{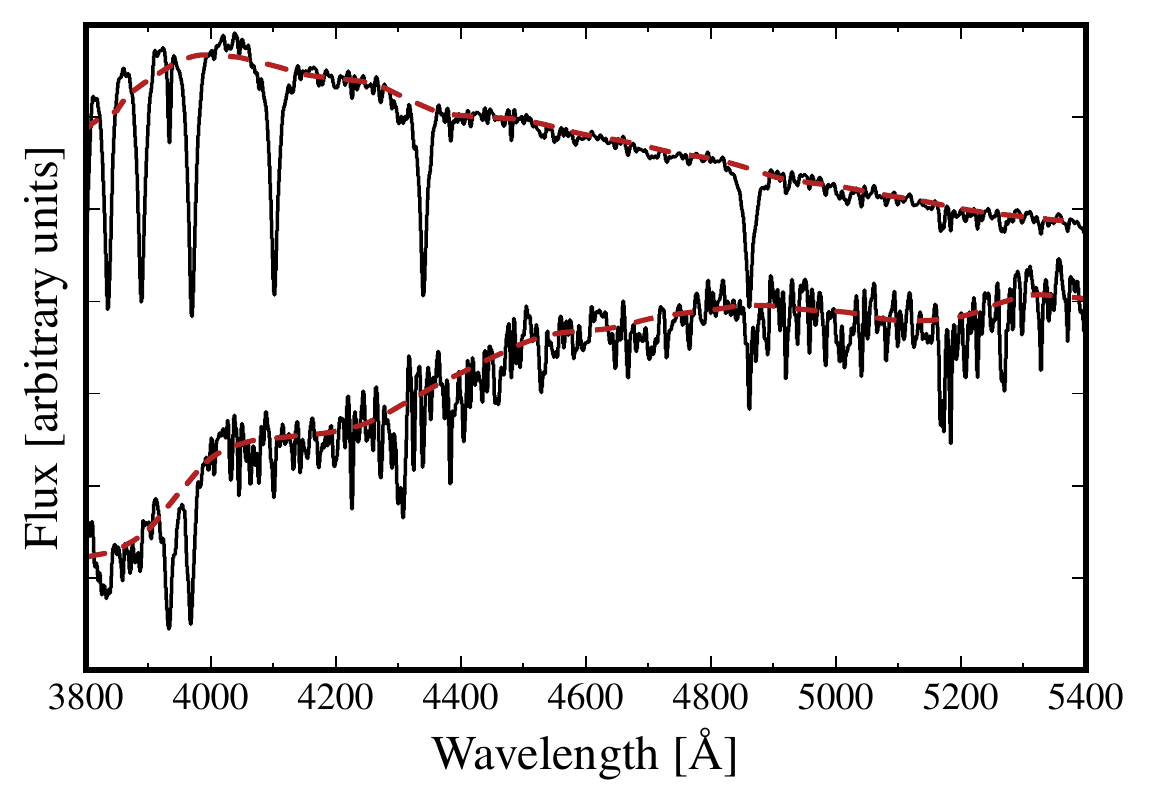}
\caption{Template spectra and their associated continuum estimates.  Dashed (red) lines show the broadband continuum estimated as described in Section \ref{py}.  Spectra are offset in flux for clarity.}
\label{fig:figure1}
\end{figure}

We follow \citet{quintero2004} and \citet{yan2006} in modelling galaxy spectra as the simple linear combination of two stellar populations, one young and one old.  Our underlying old template is a 7 Gyr old solar metallicity single stellar population (SSP) computed using the {\sc galaxev} stellar population synthesis (SPS) code \citep[hereafter BC03]{bruzual2003}.  The template was chosen to match the median luminosity-weighted age of galaxies in the SDSS \citep{gallazzi2005}, though our results are relatively insensitive to the particular choice of SSP template age.  A young population is taken 300 Myr after a starburst lasting 100~Myr, which matches the simulated duration of merger-induced starbursts \citep[e.g.][]{snyder2011}.  Our young template is selected to identify the strong Balmer absorption associated with A stars, and is therefore most sensitive to stellar populations on the order of $\sim$0.5--1.5~Gyr old.  

Prior to fitting, model spectra are broadened to match the SDSS resolution, and all spectra (templates and observed) are linearly rebinned to a dispersion of 1\AA~pixel$^{-1}$.  To avoid potential mismatches in the continuum shape between observed and template spectra we remove the broadband continuum using a moving 200\AA~wide sigma-clipped window.  In order to visually match the steep spectral shape of our burst template blue-ward of the 4000\AA~break we adopt a slightly boosted 65th percentile flux (as opposed to a true median) for determining the continuum level.  The resulting continuum is then smoothed using a 100\AA~boxcar filter; examples of the continuum fits to our template spectra are show in Figure \ref{fig:figure1}.   Emission lines are masked using an adaptive window that scales with the velocity dispersion of the galaxy\footnote{The standard SDSS {\tt spectro1d} reduction only produces velocity dispersion estimates for galaxies with an early-type spectral classification ({\tt eClass < 0}).  As such, we adopt velocity dispersion measurements from the Princeton {\tt specBS} pipeline, accessed at \url{http://das.sdss.org/}.  In general the agreement between {\tt spectro1d} and {\tt specBS} is excellent.  We refer the reader to \url{http://www.sdss.org/dr7/algorithms/veldisp.html} for details.}.  Finally, we remove any pixels in the observed spectrum with non-zero values in the SDSS bad pixel mask: spectra with fewer than 800 unmasked pixels (i.e. fewer than half in our wavelength window) are excluded from the fitting procedure altogether.

Decompositions are performed over the wavelength range 3800--5400\AA~using a non-negative least squares (NNLS) routine, and are therefore constrained to physical solutions (i.e. non-negative spectral components).  The decomposition routine returns the weight (by mass) associated with each of the young and old templates, $w_\mathrm{young}$ and $w_\mathrm{old}$ respectively; an estimate of the mass-weighted young stellar fraction, \fa, is therefore given by $w_\mathrm{young}/(w_\mathrm{young}+w_\mathrm{old})$.  Approximately 2 per cent of galaxy spectra either contain too few unmasked pixels or the NNLS algorithm fails to converge on a solution, and these objects are necessarily excluded from subsequent discussions.  We find that the vast majority of galaxies are well described by our simple two-component spectral model, with typical reduced-$\chi^2$ values of $\sim$1.1, and we find no evidence for significant variation in the quality of decompositions (in terms of $\chi^2$) with either stellar mass or derived \fa.

\begin{figure*}
\centering
\includegraphics[scale=0.88]{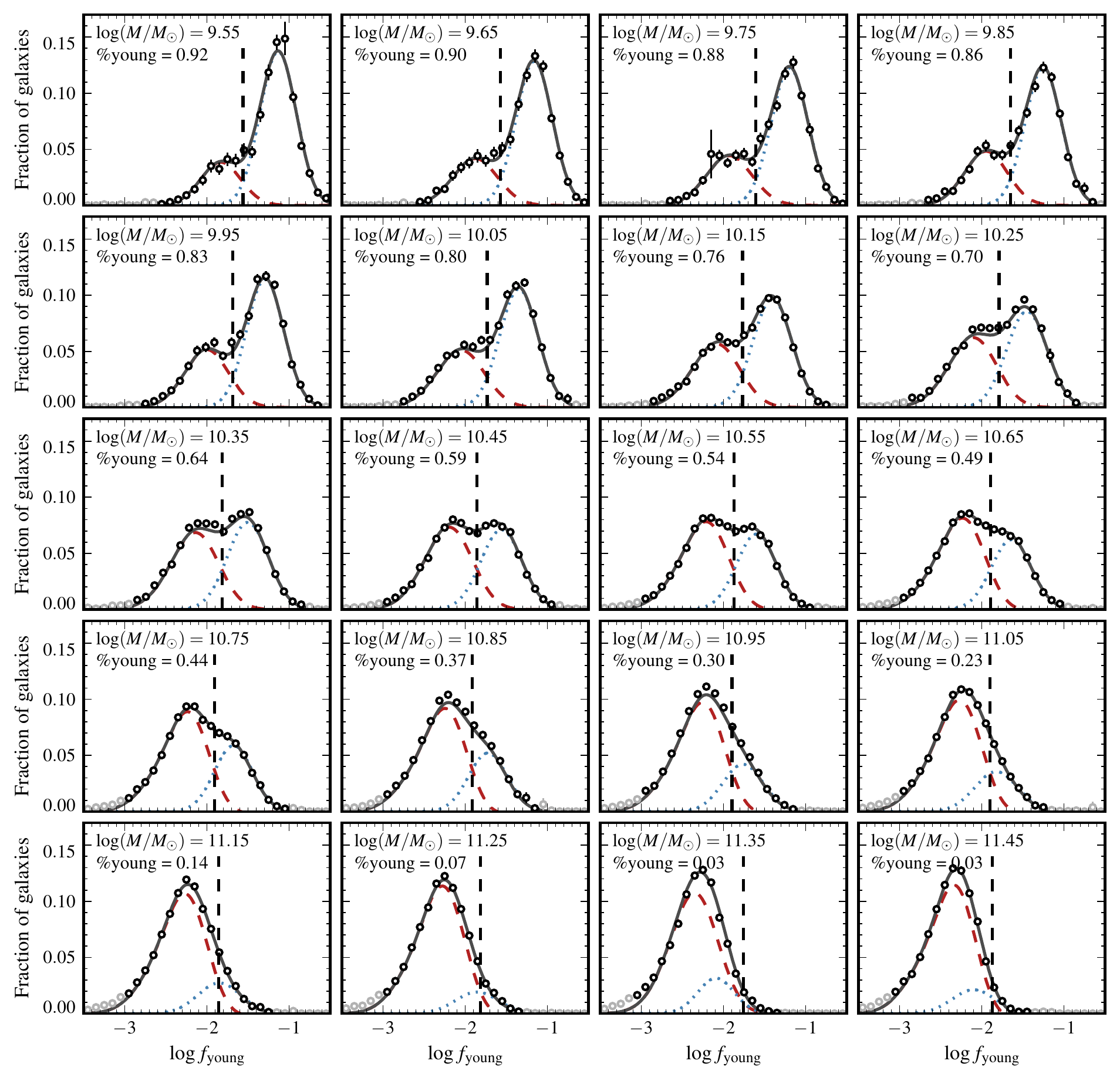}
\caption{Distribution of \fa in 0.1 dex wide bins of stellar mass for galaxies in our spectroscopic sample, where the mean mass is given in the upper left of each panel.  Open circles show the observed, $V_\mathrm{max}^{-1}$ corrected distribution of \fa, while the solid, dashed and dotted curves show the best fitting two-component Gaussian.  The vertical dashed line indicates where the ``younger'' (i.e. higher \fa) Gaussian dominates, above which we classify galaxies as young; see Section \ref{py} for details.  Light circles indicate points in the tails of the distribution that are excluded from fits.  Uncertainties on the observed data are computed based on galaxies' individual $V_\mathrm{max}^{-1}$ measurements added in quadrature.}
\label{fig:figure2}
\end{figure*}

Star-formation histories are known to vary systematically as a function of galaxies' stellar mass, such that massive galaxies are older and form stars more rapidly than their low-mass counterparts.  In the context of identifying quenched galaxies, this leads us to adopt a relative definition of what makes a galaxy ``young'' so that we can account for these systematic variations in our selection.  In Figure \ref{fig:figure2} we show the distribution of \fa in bins of stellar mass for galaxies with $9.5 \leq \log(M/M_\odot) \leq 11.5$.  We see that both the mean \fa and the overall fraction of young galaxies decrease with increasing stellar mass, as we might expect from the (known) correlation between stellar mass and age.  In each bin we model the \fa distribution as the sum of two Gaussians (dashed and dotted lines in Figure \ref{fig:figure2}).  In order to limit the influence of modelling uncertainties on our classification we exclude galaxies with extreme values of \fa (shown as light points in Figure \ref{fig:figure2}).  The vertical dashed line show the point where the ``younger'' (i.e., higher \fa) Gaussian dominates, and above this point we classify galaxies as young.  In general, our selection of young galaxies is limited to those with \fa greater than 1 to 2 per cent by mass (i.e. \fa $\sim$ 0.01--0.02).

\subsubsection{Reliability of \fa}
\label{reliability}

\begin{figure}
\centering
\includegraphics[scale=0.69]{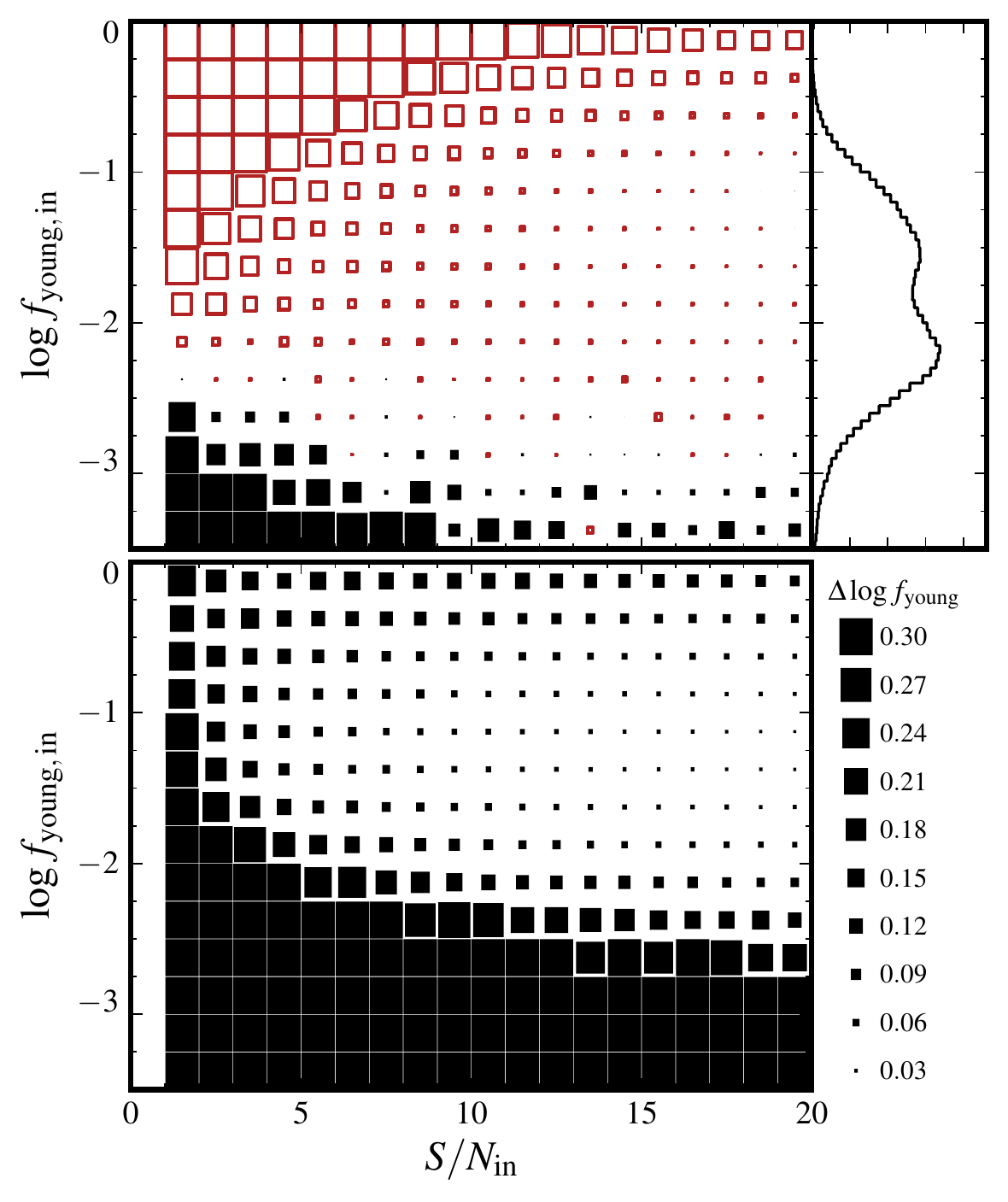}
\caption{Results of the Monte Carlo simulations described in Section \ref{reliability} for the recovery of \fa.  In the top left panel we show the systematic offset in \fa as a function of input \fa and \sn, where cell size scales with the magnitude of the offset (as indicated) up to a maximum factor of 2 uncertainty.  Open (red) and filled (black) squares indicate negative and positive offsets, respectively.  For comparison, the top right panel shows the observed distribution of \fa.  In the bottom panel we show the statistical uncertainty as a function of \fa and \sn, where cells are scaled as above.}
\label{fig:figure3}
\end{figure}

Before proceeding it is useful to assess the overall reliability of our \fa measurements as a function of both spectral \sn and \fa.  To this end, we have performed a set of Monte Carlo simulations where we randomly generate composite spectra with a range of \fa and \sn, and then process these synthetic data using the procedures outlined above.  We show the results of these simulations in Figure \ref{fig:figure3}, where the size of individual cells indicates either the systematic (top panel) or statistical (bottom panel) uncertainty in the measured \fa as a function of simulated (true) \fa and \sn.  As one might expect there is a strong dependence of the uncertainty in \fa as a function of \sn, such that parameters are significantly unreliable---both systematically and statistically---at low \sn.  Relative to the observed distribution of \fa (top right panel of Figure \ref{fig:figure3}), significant systematic offsets are only avoided for spectra with $S/N \geq 10$, which leads us to adopt this as a \sn requirement for the remainder of this work.

\begin{figure}
\centering
\includegraphics[scale=0.69]{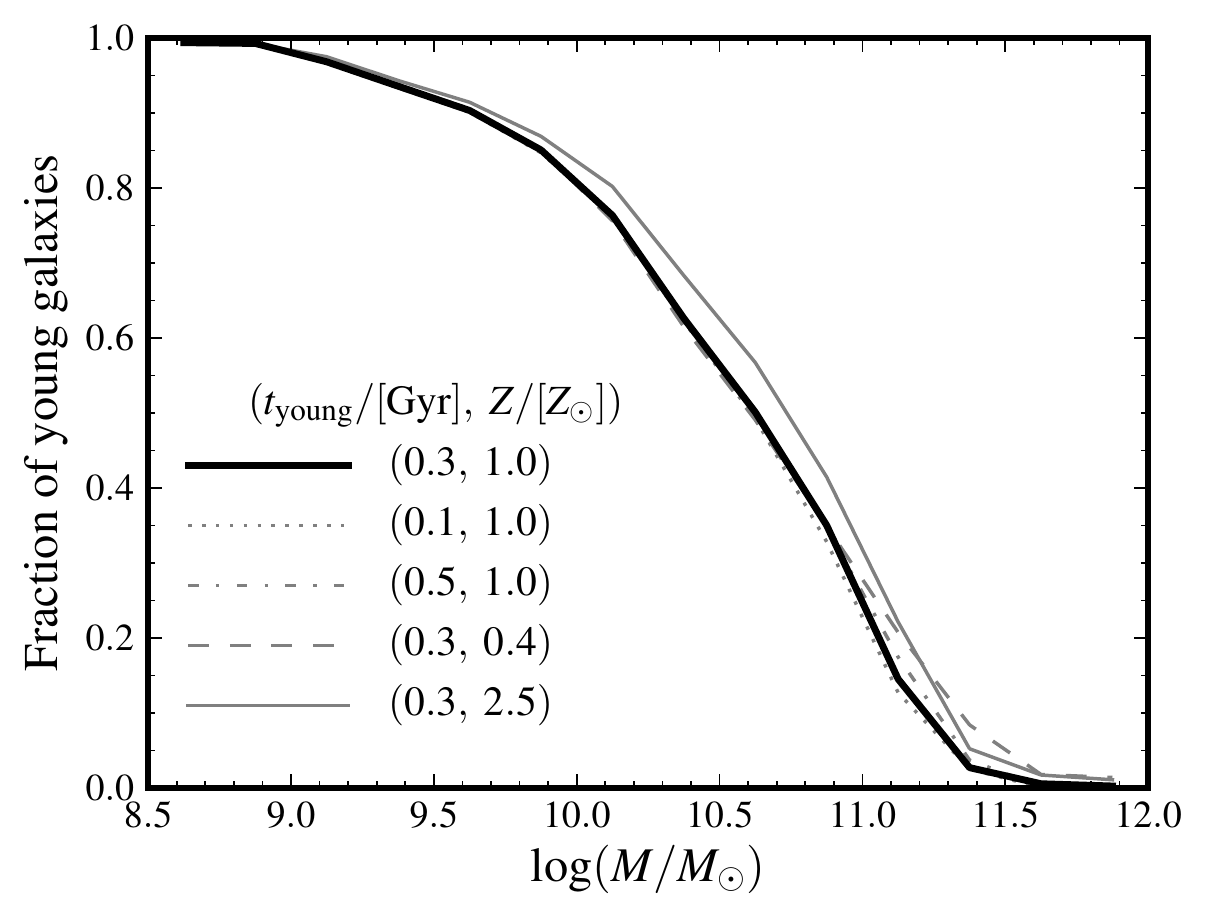}
\caption{Fraction of young galaxies as a function of stellar mass for different spectral templates.  The thick solid line shows our fiducial decomposition, where the burst is 300 Myr old and both components have solar metallicity.  Dotted, dot-dashed, dashed and solid lines show the young fractions derived for variation in the burst template age of $\pm$200 Myr or the metallicity of both templates by a factor of 2.5.}
\label{fig:figure4}
\end{figure}

Alternatively, we can ask about the sensitivity of our derived results to the parameters used in our spectral decompositions.  In particular, we would like to know if (and how) our choice of burst template age (300 Myr) or metallicity ($Z = Z_\odot$) may bias our overall classifications of galaxies as young.  In Figure \ref{fig:figure4} we plot the derived fraction of young galaxies as a function of stellar mass, where young galaxies are classified as described in Section \ref{py}.  The thick solid line shows the fraction of young galaxies as a function of stellar mass derived using our fiducial templates.  For comparison, the thin dotted, dashed, dot-dashed and solid lines show the fraction of young galaxies determined from fits in which either the burst template age or overall metallicities are varied as indicated in the Figure.  We find that, while changing the template ages or metallicities leads to a factor of $\sim$2 variation in the \emph{absolute} value of \fa---that is, the inferred mass fraction of young stars---the overall \emph{fraction} of galaxies classified as young is relatively insensitive to variations in the underlying templates due to our relative (as opposed to absolute) qualification of ``young''.

Finally, it is important to note that the fixed 3$^{\prime\prime}$ aperture of the SDSS fibres leads to a variation in physical scale of $\sim$0.6 to 10 kpc over the redshift range of our sample.  We must therefore be concerned that any radial variation in galaxies' age or metallicity could lead to a systematic bias in our measurements of \fa.  We search for any such dependence in by examining the relationship between \fa and $z/z_\mathrm{max}$, where $z_\mathrm{max}$ is the maximum redshift at which a galaxy could be included in our sample, discussed previously in Section \ref{sampling}.  Our data show evidence for an increase in \fa with increasing redshift, although any such variation in passive galaxies is relatively small ($<$1 per cent across the volume of the sample) and does not appear to significantly affect our spectral classifications.  The observed increase in \fa is consistent with an increasing covering fraction at higher redshift coupled with the negative colour gradients observed for many early-type galaxies \citep[e.g.][]{peletier1990,goudfrooij1994,roche2010}.  While it does not appear that aperture effects significantly influence our results, we nevertheless reiterate that our stellar population measurements are restricted to the central few kiloparsecs in most galaxies, and hence do not necessarily reflect their global properties.

\subsection{Quenched galaxy sample}
\label{spec_prop}

Our final sample of quenched galaxies comprises all passive galaxies, as determined by their SSFR, that show evidence for a young stellar population, defined as above.  In addition, we limit our sample to galaxies with \logm$\geq 9.5$.  All told, we identify 12,105 such systems, 2.5 per cent of our parent SDSS sample, with a space density of $\sim$$2.2\times10^{-4}$~Mpc$^{-3}$.

\begin{figure}
\centering
\includegraphics[scale=0.65]{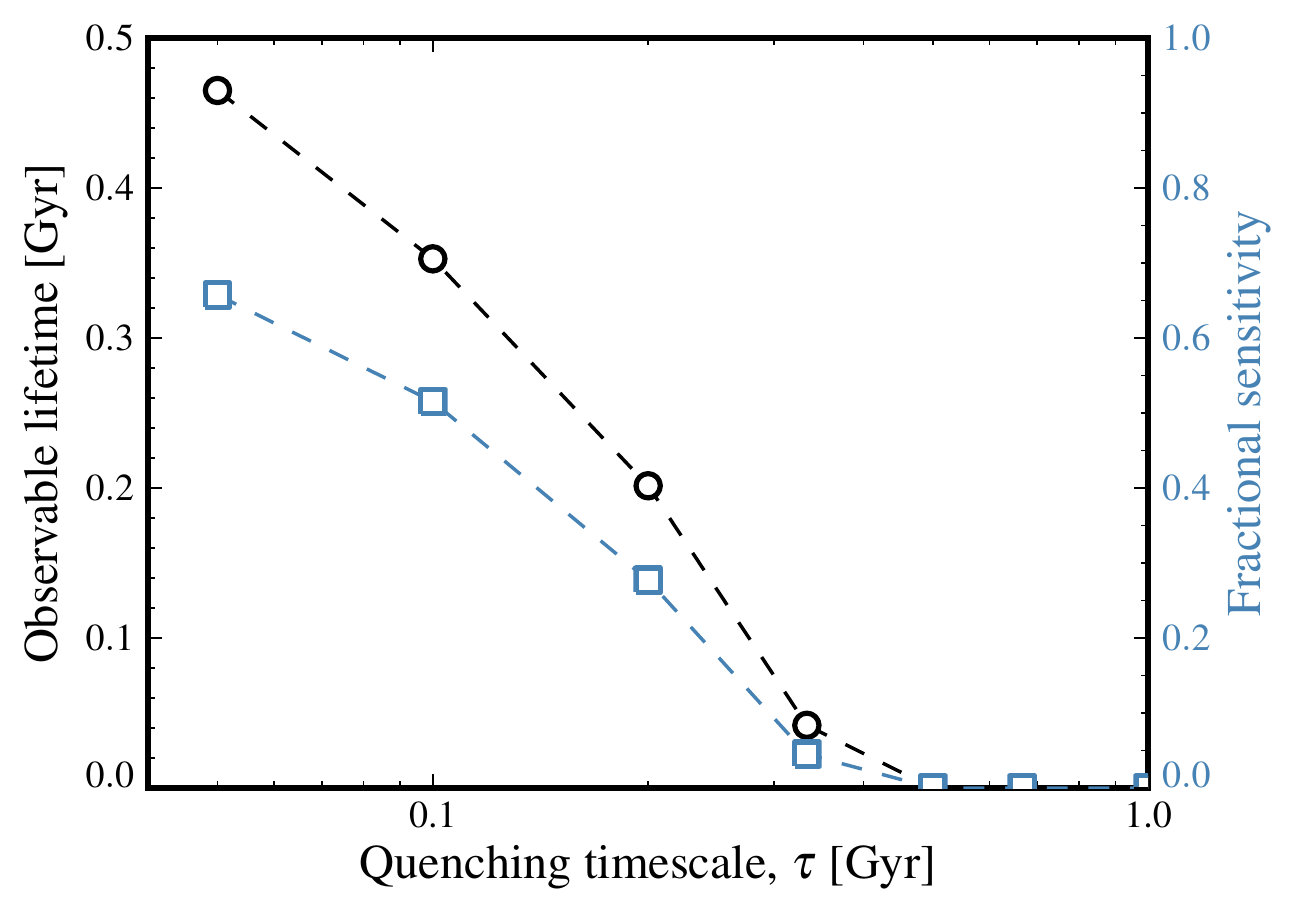}
\caption{Observability of simulated quenched galaxies as a function of the quenching rate, characterised by the $e$-folding time, $\tau$.  The left axis and open (black) circles show the detectability of quenched galaxies given our selection criteria (Section \ref{spec_prop}) for values of the decay rate.  The right axis and open (blue) squares indicate the fraction of their \emph{total} transition lifetime---defined here as the time spent with $1.4 \leq \mathrm{D_n4000} \leq 1.7$---for which simulated galaxies are detectable as quenched.}
\label{fig:figure5}
\end{figure}

The primary factor governing our selection of quenched galaxies is the rate at which star formation is truncated.  In order to illustrate this point,  we construct a suite of synthetic star-formation histories described by 8 Gyrs of constant star formation followed by an exponential decline; by varying the characteristic decay rate of the decline, $\tau$, we can simulate a wide variety of quenching timescales. Such models have been shown to provide a relatively good match to observed properties of the star-forming galaxy population \citep[e.g.][]{brinchmann2004}.  In Figure \ref{fig:figure5} we plot the timescale over which galaxies would be observable given our quenched galaxy selection criteria---in this case, the timescale for which a given model has \fa > 0.01 and $\log \mathrm{SSFR} < -11$---as a function of $\tau$.  This highlights the clear dependence of our sample selection on quenching timescale: galaxies whose star formation stops abruptly are readily identified as quenched, while galaxies whose star formation declines over $\gtrsim$Gyr timescales are likely not included in our quenched galaxy sample.  Alternatively, we can cast these simulations in terms of their detectability by scaling their observable lifetimes according to the \emph{total} time that quenched galaxies spend in transition between the star-forming and passive populations.  Here we adopt a relatively simple definition for galaxies' transition lifetime based on the time that simulated spectra spend with $1.4 \leq \mathrm{D_n4000} \leq 1.7$, i.e. with intermediate spectral properties.  The results of this scaling are indicated on the right-hand axis of Figure \ref{fig:figure5} by open squares.  For quenching timescales of $\tau \lesssim 0.5$~Gyr we find that galaxies are classifiable as quenched for $\sim$10 to 60 per cent of their transition lifetimes, depending on $\tau$.

The extent to which our sample is representative of the ``true'' transition population therefore depends sensitively on the typical timescales over which star formation declines.  \citet{kauffmann2004} argue that the lack of a clear environmental dependence in the joint distribution of \hda and \dfk points towards slower timescales for galaxies' transition from star forming to passive, $\gtrsim$1 Gyr, in agreement with the median timescale of $\sim$1.5 Gyrs derived by \citet{kaviraj2007} based on the evolution of $UV$ and optical colours.  However, a slow evolution for \emph{most} galaxies seems to be at odds with the persistent bimodal structure of both the SSFR and colour distributions: if most galaxies transition from star-forming to passive over time scales $\gtrsim$1-1.5 Gyrs, then such galaxies should fill in the apparent valley at intermediate SSFR and colour \citep[e.g.][]{balogh2004,wetzel2012}.  Of course, the most likely scenario is that there is no \emph{single} $\tau$ that describes galaxies' transition from star-forming to passive, and therefore it is more instructive to consider the distribution of $\tau$, as in \citet{kaviraj2007} and \citet{martin2007}.  \citet{martin2007} use a combination of $UV$--optical colour and absorption indices (\hda and \dfk) to show that the majority of transition galaxies at low redshift ($\sim$60--80 per cent) are best described by models with $\tau \lesssim$ 0.5--1 Gyr once selection effects are accounted for, and that a relative minority of galaxies are found with $\tau > 1$ Gyr.  Based on Figure \ref{fig:figure5}, we would therefore expect to identify a majority of quenched galaxies with our adopted selection criteria ($\gtrsim$50 per cent), with the caveat that our results are biased towards those galaxies with the most rapid truncation timescales, e.g. $\tau <$ 0.2--0.3 Gyrs.

\begin{figure}
\centering
\includegraphics[scale=0.69]{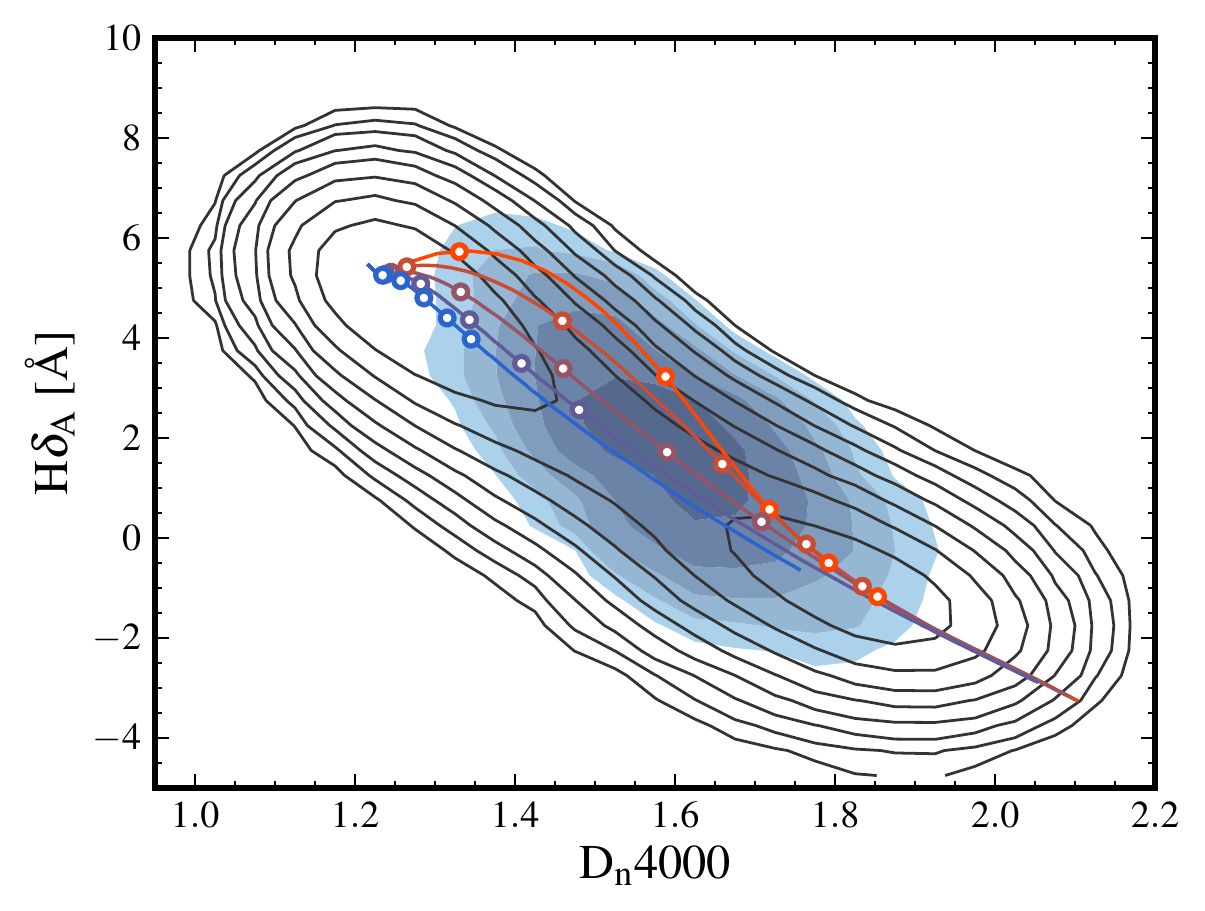}
\caption{The joint distribution of \hda and \dfk for all galaxies in our SDSS sample (black contours) and for the spectroscopically-selected quenched galaxies (shaded contours).  Lines show the predicted evolution of quenched galaxies following 8 Gyrs of constant star formation, where different colours (orange to blue) indicate different exponential decay rates of $\tau$ = 2, 1, 0.5, 0.2 and 0.05 Gyrs. Open circles show indicate 0.1, 0.5, 1.0, 1.5 and 2.0~Gyrs after the onset of quenching.}
\label{fig:figure6}
\end{figure}

\begin{figure}
\centering
\includegraphics[scale=0.88]{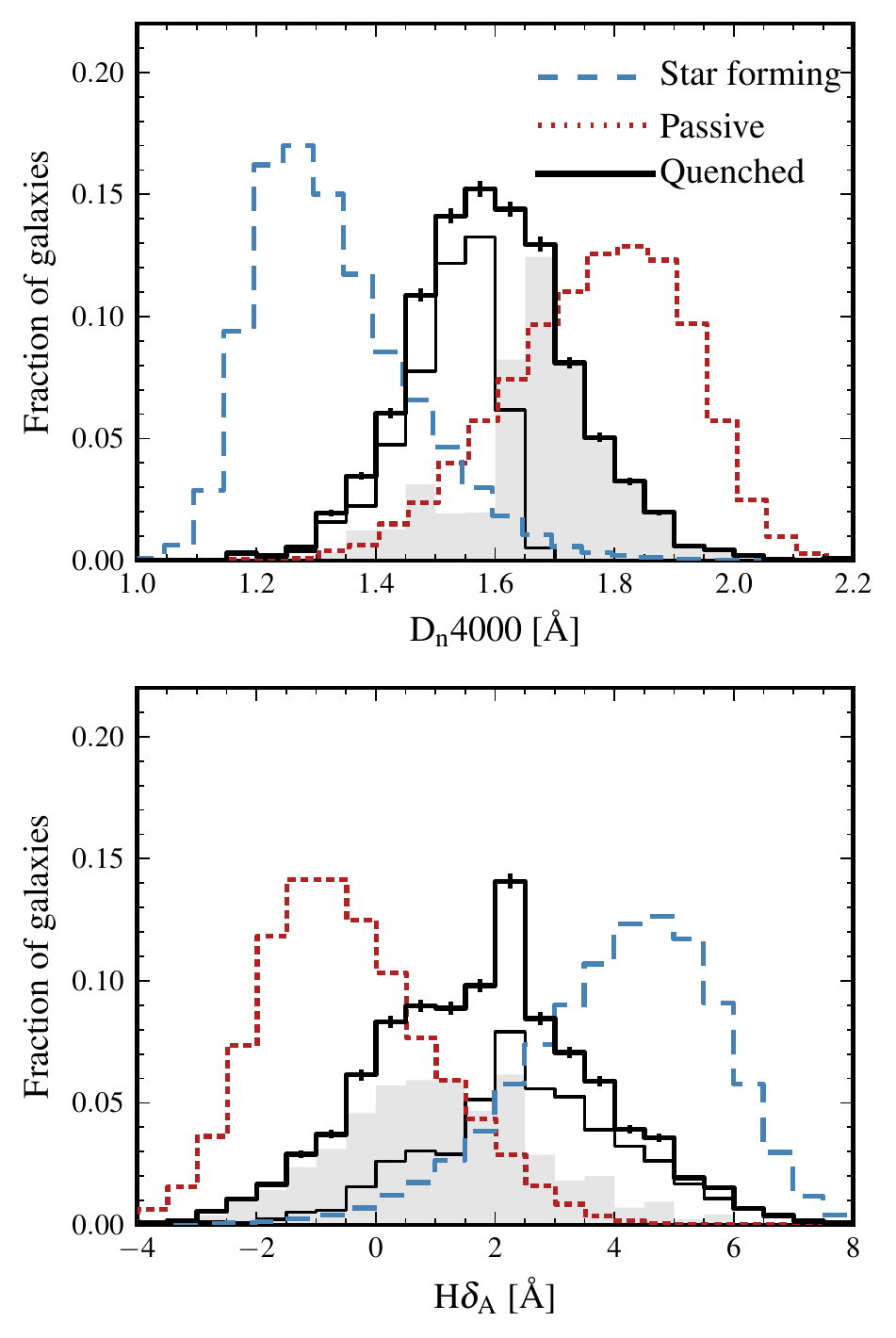}
\caption{Distribution of \dfk (\emph{top panel}) and \hda (\emph{bottom panel}) for star-forming, passive and quenched galaxies (dashed, dotted and solid lines, respectively).  Quenched galaxies are intermediate systems in terms of both their \dfk and \hda values.  The shaded histogram shows the distribution of \dfk and \hda for quenched galaxies that satisfy our SSFR cut, while the thin, open histogram shows the distribution for galaxies with a high SSFR from \citetalias{brinchmann2004}, but low EW(\ha) ($\geq -3$~\AA; see Section \ref{quenched_select}).}
\label{fig:figure7}
\end{figure}

We can obtain a more quantitative picture of our quenched galaxy sample by considering their spectroscopic properties relative to the general star forming and passive populations.  As discussed previously, H$\delta$ absorption is primarily sensitive to stars with intermediate spectral types (A through early F), and therefore traces a particular mass range of stars as they evolve over $\sim$0.5--1.5~Gyr timescales.  By comparison, the spectral break at 4000\AA~occurs as a result of metal-line blanketing in the atmospheres of cool stars blue-ward of 4000\AA; in hot stars these metal species are multiply ionised, and therefore the strength of the 4000\AA~break traces the overall ageing of a stellar population as progressively cooler stars dominate the stellar light.  Considering galaxies in the parameter space defined by their 4000\AA-break strength and H$\delta$ absorption therefore provides a useful diagnostic for understanding the evolution of young stellar populations, and has been used by many authors in the past for just this purpose \citep[e.g.][]{kauffmann2003b,kauffmann2004,wild2007,von-der-linden2010}.  In Figure \ref{fig:figure6} we plot the distribution of \hda \citep{worthey1997} against a \dfk for the galaxies in our sample.  Open contours show the overall distribution of the SDSS sample, while the shaded contours show the parameter space occupied by quenched galaxies.  The distribution of \dfk and \hda is roughly bimodal, where star-forming galaxies occupy a locus with high \hda and low \dfk (i.e. the upper-left corner of Figure \ref{fig:figure6}), while old passive galaxies have low \hda and high \dfk.

The distribution of quenched galaxies in Figure \ref{fig:figure6} can be understood via comparison with the expected behaviour of the truncated star-formation models discussed previously, which are shown in Figure \ref{fig:figure6} for a variety of $e$-folding timescales as solid curves.  In the case of a near-instantaneous truncation of star-formation (orange line in Figure \ref{fig:figure6}), \hda increases as O and B stars terminate their evolution and the observed spectrum becomes dominated by A stars.  The subsequent steady decline of \hda and increase of \dfk follows the evolution of the stellar population as it ages.  Galaxies with the highest values of \fa correspond to the strongest Balmer-line absorbers (i.e. the upper left of Figure \ref{fig:figure6}), and therefore represent the subset of quenched systems caught early in their evolution with rapid truncation times (i.e. smallest $\tau$).  A fraction quenched galaxies show extreme values of \hda that cannot be accounted for in the simple truncated star-formation histories considered here, and are more readily described by recent bursts of star-formation; these galaxies can be though of as ``classical'' post-starbursts, and make up a relative minority of the sample considered here.
 
A more direct comparison between the properties of quenched galaxies and the general star-forming and passive galaxy populations is shown in Figure \ref{fig:figure7}, where we plot the distributions of \hda and \dfk separately.  In both panels, histograms are individually normalised to unity so that the $y$-axis represents the fraction of galaxies in a given population.  In terms of their \dfk distribution, quenched galaxies form a well defined peak, owing to the tight relationship between \dfk and luminosity-weighted age.  We also show the distribution of galaxies selected as passive based purely on the SSFRs of \citetalias{brinchmann2004} and those classified as passive based on their EW(\ha) as shaded and open (thin) histograms, respectively.  There is a clear division between the two selections at \dfk $\sim$ 1.6, which is a direct result of our adopted SSFR limit of $\sim$10$^{-2}$~Gyr$^{-1}$ and the relationship between \dfk and SSFR adopted by \citetalias{brinchmann2004} (their figure 11).

The distribution of \hda in quenched galaxies is generally broader than that of \dfk, in part due to the large uncertainty of the \hda index relative to \dfk \citep[cf. figures 5 and 6 of][]{wild2007}.  Nevertheless, quenched systems occupy a relatively distinct peak of \hda intermediate between passive and star-forming galaxies.

\section{The physical properties of quenched galaxies}
\label{results}

The criteria outlined in Section \ref{quenched_select} aim to select those galaxies that have recently joined the passive population.  In this section, we examine the colour, morphology and emission-line (AGN) properties of quenched galaxies in order to place constraints on their progenitors and identify plausible mechanisms associated with their transition.

\subsection{Global colours}
\label{colour}

\begin{figure}
\centering
\includegraphics[scale=0.88]{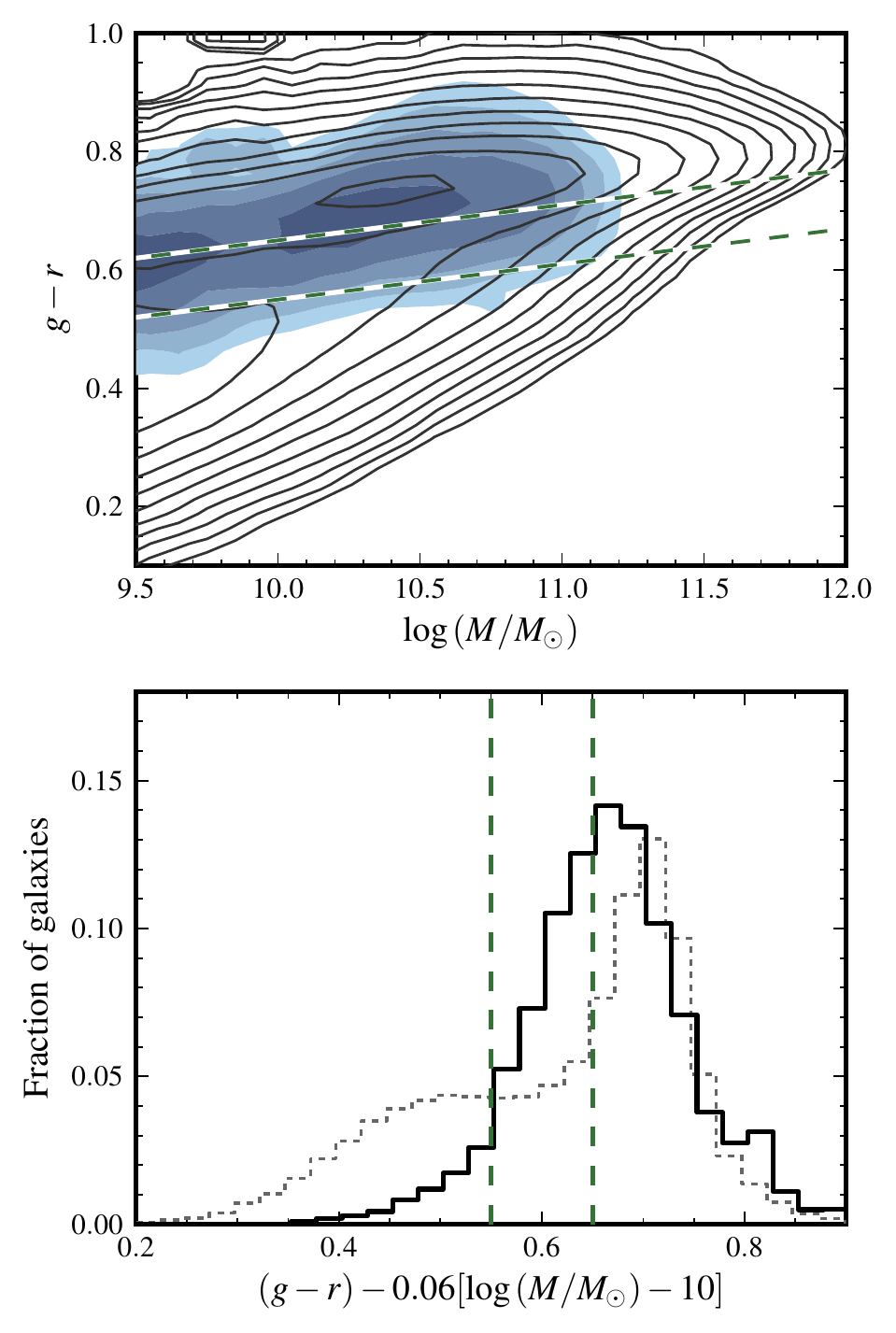}
\caption{The colour--stellar mass relation (top panel) and colour distribution (bottom panel) for galaxies in our spectroscopic sample with \logm $\geq 9.5$.  In the top panel, contours show the colour--stellar mass relation for our full sample, while shading shows the distribution of quenched galaxies.  The bottom panel shows the rest-frame $g-r$ colour distributions for all (thin line) and quenched (thick line) galaxies, taking account for the slope in the CMR.  In both panels the dashed lines bracket the region identified as the `green valley' (see Section \ref{colour} for details).}
\label{fig:figure8}
\end{figure}

As a starting point, we consider the relationship between galaxy colour and stellar mass.  As discussed in Section \ref{intro}, the colour--mass relation (CMR) is bimodal, with two separate but parallel sequences representing quiescent and star-forming galaxies.  In the upper panel of Figure \ref{fig:figure8} we show the CMR for galaxies in our full sample (black contours); the region of the colour--mass plane occupied by quenched galaxies is indicated by shaded contours.  A significant fraction of quenched galaxies lie intermediate between the blue cloud and red sequence---dubbed the `green valley'---with a tendency to be found near the red sequence.  This can be seen more clearly in the bottom panel of Figure \ref{fig:figure8}, where we plot the $g-r$ colour distribution both for galaxies in the full sample (dotted histogram) and only quenched galaxies (solid histogram).

We can gain a better understanding of how galaxies are distributed in the colour--mass plane using a simple form for the mass-dependent division between red, green and blue galaxies of

\begin{equation}
g-r = A + 0.06[\log\,(M/M_\odot)-10],
\end{equation}

\noindent where 0.06 sets the slope of the CMR.  We follow \citet{mendez2011} and \citet{lackner2012} in adopting a width for the green valley of 0.1 mags in $g-r$, centred on the minimum of the CMR at \logm = 10 ($A = 0.60$; shown by the green lines in Figure \ref{fig:figure8}); red and blue galaxies are defined as those above and below the green valley, respectively.  

In terms of quenched galaxies 57, 36 and 7 per cent are classified as red, green and blue, respectively, compared to 52, 19 and 29 per cent for the full SDSS sample.  While it is relatively easy to understand the dearth of blue galaxies in our quenched sample, given the prevalence of star formation in the blue cloud population, the relatively high abundance of red galaxies may be somewhat surprising.  To a large extent this is a result of galaxies' non-linear transition from blue to red following the truncation of star formation, which leads to galaxies evolving rapidly through intermediate colours and piling up as they approach the red sequence.  It may also be that our selection criteria are biased against including the youngest (bluest) systems.  For example, if the majority of quenching occurs as a result of AGN activity, then the youngest quenched galaxies (with \dfk $<1.6$) may still have significant \ha emission \citep[e.g.][]{schawinski2007}, prohibiting their inclusion in our sample.  However, it is also important to stress that the \emph{majority} of green galaxies ($\sim$80 per cent) are classified as star-forming according to the criteria outlined in Section \ref{quenched_select}, suggesting a clear distinction between colour- and spectroscopically-selected samples \citep[see also][]{woo2012}.

The green valley is generally assumed to host galaxies for a short period of time as they transition from star forming to passive \citep[e.g.][]{martin2007,mendez2011}.  Indeed, as discussed earlier, galaxies' colour \emph{must} evolve rapidly in order to preserve the relative under-abundance of green galaxies relative to blue or red \citep[e.g.][]{balogh2004}.  We can investigate this scenario explicitly with our data by comparing the relative space densities of quenched and green galaxies; although our sample is biased towards the later stages of quenching (i.e. towards red colours), the relative abundance of green and quenched galaxies should be comparable if there is no significant evolution in the quenching rate on $\sim$Gyr timescales.  Quenched galaxies are a factor of $\sim$6 under-abundant compared with a colour-selected sample, with a space density of $2.2\times10^{-4}$~Mpc$^{-3}$ versus $1.5\times10^{-3}$~Mpc$^{-3}$ for green galaxies.  However, as discussed in Section \ref{spec_prop}, our selection criteria are predominantly sensitive to galaxies experiencing relatively rapid quenching.  Figure \ref{fig:figure5} suggests that we may miss over half of quenched galaxies with $\tau < 0.5$ due to their short observational lifetimes, along with a further 50 per cent of galaxies whose star formation is quenched over relatively longer timescales.  Adopting this factor of $\sim$4 suggests that up to 60 per cent of present-day green valley galaxies will evolve onto the red sequence over $\sim$Gyr timescales, while the remaining $\sim$40 per cent may represent dusty star-forming galaxies rather than true transition objects.

\subsection{Visual and quantitative morphology}
\label{vmorph}

\begin{figure}
\centering
\includegraphics[scale=0.88]{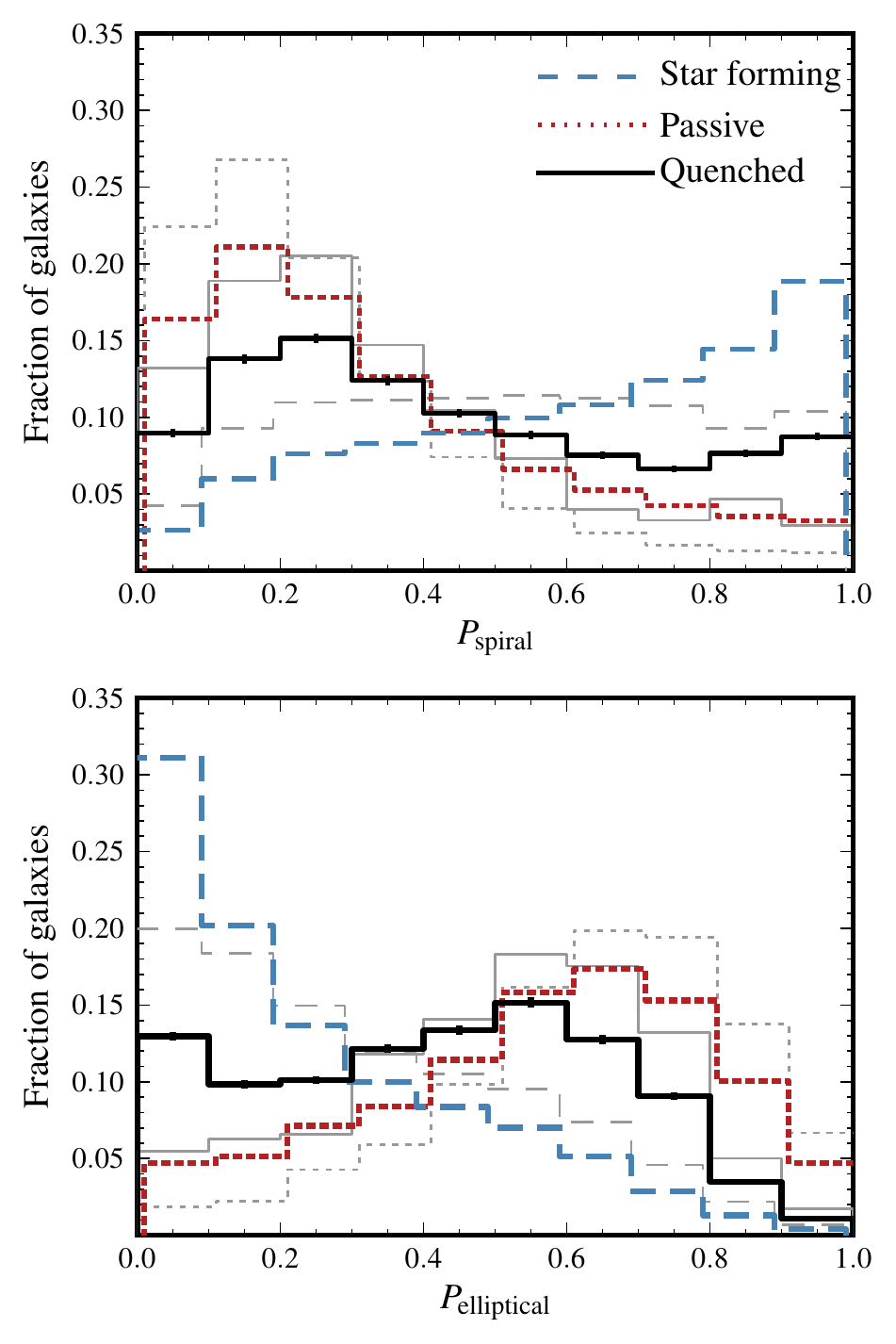}
\caption{Distribution of spiral (top panel) and elliptical (bottom panel) morphological probabilities for different galaxy samples.  Star-forming and passive galaxies (dashed and dotted histograms, respectively) are defined using the division in SSFR discussed in Section \ref{quenched_select}, while quenched galaxies (solid line) shows passive galaxies weighted by \py.  Thick (coloured) lines show the distribution of morphological probabilities for all galaxies, while the grey histograms show the distributions for galaxies with $(b/a) < 0.5$ ($i < 60^\circ$).}
\label{fig:figure9}
\end{figure}

\begin{figure}
\centering
\includegraphics[scale=0.88]{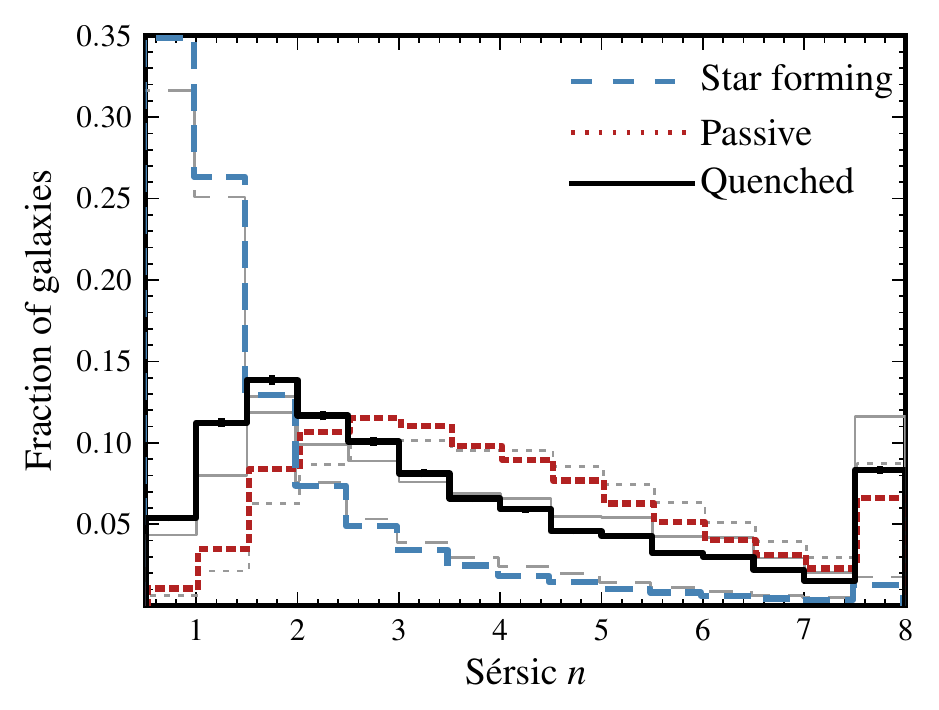}
\caption{Distribution of S\'ersic indices for star-forming, passive and quenched galaxies (dashed, dotted and solid histograms, respectively).  Thick (coloured) lines show the distribution of morphological probabilities for all galaxies, while the grey histograms show the distributions for galaxies with $(b/a) < 0.5$ ($i < 60^\circ$).}
\label{fig:figure10}
\end{figure}

Different physical scenarios make different predictions for how star formation is quenched; however, based on purely phenomenological arguments, the transition of galaxies from star forming to passive must, at some level, carry with it an associated transition from disk dominated to spheroid dominated in order to jointly preserve the morphology--density and star-formation rate--density relations.  Examining the morphological properties of quenched galaxies therefore offers an important constraint on the relative timescales for morphological transformation and star-formation truncation.

In Figure \ref{fig:figure9} we plot the distributions of Galaxy Zoo elliptical and spiral probabilities for star-forming, passive and quenched galaxies.  This figure shows that quenched galaxies are intermediate between star-forming and passive galaxies in terms of their visual appearance, with relatively few confidently classified as spiral or elliptical.   In addition, there is an excess of quenched galaxies with low elliptical probability (or, alternatively, high $P_\mathrm{spiral}$) relative to other passive galaxies.  This excess is consistent with an increased fraction of quenched galaxies hosting both bulges and disks, supported by the fact that the disparity between passive and quenched galaxies is reduced by considering only face-on galaxies ($b/a < 0.5$; thin grey lines in Figure \ref{fig:figure9}).

We can better characterise any structural differences between the star-forming, passive and quenched subpopulations using the quantitative morphological decompositions from \citet{simard2011}; we show in Figure \ref{fig:figure10} the distribution of S\'ersic indices for these different subpopulations.  The separation between star-forming and passive galaxies in terms of their structure is clearly represented here, where star-forming galaxies are predominantly disky ($n < 1.5$) and passive galaxies are more centrally concentrated ($n > 2$).  Most strikingly, there is a significant deficit of low S\'ersic index galaxies ($n \lesssim 1.5$) in the quenched population relative to star-forming galaxies, suggesting that the hallmark of quenched galaxies is their increased central concetration.  This confirms our expectation based on the analysis of visual classifications in Figure \ref{fig:figure9}, i.e. that quenched galaxies are structurally more similar to (already) passive galaxies than star-forming ones, and is consistent with a picture where the transition of galaxies from star forming to passive (or, alternatively, blue to red) is accompanied by a change in their morphologies \citep[see also][]{mendez2011,lackner2012}.  As noted in the visual classifications, there is an increase in the fraction of quenched galaxies with $P_\mathrm{spiral} > P_\mathrm{elliptical}$ relative to the general passive population ($\sim$46 vs. 27 per cent).  This increase is mirrored by a similar rise in the fraction of galaxies with $n \leq 2.5$ ($\sim$42 vs. 24 per cent), and may suggest that many quenched galaxies host disks in addition to significant stellar bulges.

\subsection{Incidence of AGN}
\label{agn}

The relationship between star-formation and AGN activity remains a hotly debated topic.  While from a theoretical perspective AGN are thought to play a crucial role in regulating star-formation activity, observational evidence for a {\it causal} link between AGN activity and the suppression or cessation of star formation is tenuous.  More generally, while it is likely that feedback is related in some way to star-formation activity, the exact nature of that relationship remains poorly understood.  In this picture, quenched galaxies offer an attractive target for the study of feedback processes based on their discontinuous star-formation histories.
 
\begin{figure}
\centering
\includegraphics[scale=1.0]{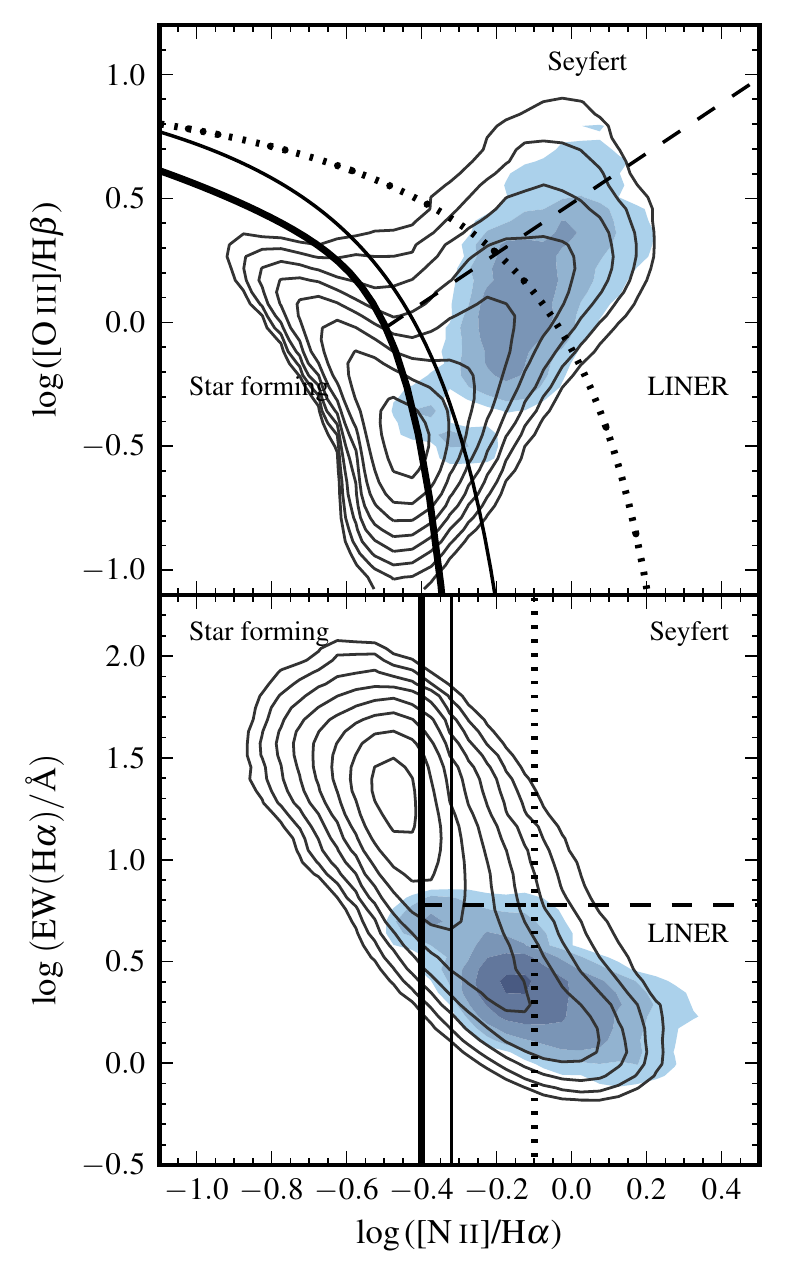}
\caption{Line ratio diagnostics for galaxies in our SDSS sample. \emph{Top panel}: BPT diagram for the 143,727 galaxies ($\sim$30 per cent) with $S/N \geq 3$ in each of \hb, \oiii, \ha and \nii shown as open contours, with quenched galaxies shown as shaded contours.  Lines show different divisions between star-forming and AGN galaxies proposed by \citetalias{stasinska2006} (thick solid line), \citetalias{kauffmann2003} (thin solid line) and \citetalias{kewley2001} (dotted line).  The dashed line shows the division between Seyfert and LINER galaxies proposed by \citet{cid-fernandes2010}.  \emph{Bottom panel}: The equivalent width \ha vs. \nii/\ha (WHAN) digram proposed by \citet{cid-fernandes2010} for the 263,530 galaxies ($\sim$54 per cent) with $S/N \geq 2$ in \ha and \nii.    Lines are the same as in the above panel, but now represent the `optimal transposition' of the \citetalias{kauffmann2003}, \citetalias{stasinska2006} and \citetalias{kewley2001} classifications into EW(\ha)--\nii/\ha parameter space as derived by \citet{cid-fernandes2010}.}
\label{fig:figure11}
\end{figure}

In general, optical classification of galaxies as either AGN hosts or star forming is based on the comparison of emission-line ratios \citep[e.g.][hereafter BPT]{baldwin1981} with either empirically-derived or theoretically-determined divisions between different classes.  In the top panel of Figure \ref{fig:figure11} we show three such demarcations from \citet[hereafter S06]{stasinska2006}, \citet[hereafter K03]{kauffmann2003} and \citet[hereafter K01]{kewley2001} as thick solid, thin solid and dotted lines, respectively, while the straight dashed line shows the proposed `optimal' division of AGN between LINERs and Seyferts determined by \citet[see also \citealp{kewley2006}]{cid-fernandes2010}.  Contours show the distribution of galaxies in our full sample with $S/N \geq 3$ in each of \hb, \oiii, \ha and \nii, while shading indicates the location of quenched galaxies satisfying the same \sn limits.  Perhaps unsurprisingly, the majority of quenched galaxies with detectable emission lines are classified as AGN (99.3, 98.3 and 41.8 per cent for \citetalias{stasinska2006}, \citetalias{kauffmann2003} and \citetalias{kewley2001} classifications, respectively), and the vast majority of these have LINER-like spectra ($\sim$60--80 per cent depending on classification).  As discussed in Section \ref{quenched_select}, the abundance of LINER-like emission in quenched galaxies may not be surprising.  While our selection of galaxies based on the combination of SSFR and EW(\ha) is more inclusive than, say, a cut on \emph{just} \ha or \oii \citep[e.g.][]{yan2006,yan2009}, we are unable to properly account for the population of strong-line AGN.  As an added complication, requiring detection in four emission lines limits the extent to which weak-lined systems are even classifiable under the BPT scheme as both \hb and \oiii are fainter and harder to detect than \ha and \nii.

As an alternative to the BPT diagram, \citet{cid-fernandes2010} propose a weak-line classification scheme that uses EW(\ha) in place of \oiii/\hb---the EWH$\alpha$n2 or WHAN \citep{cid-fernandes2011} diagram---allowing the classification of a significant number of galaxies with measurable \ha and \nii flux, but poor detections of the weaker \hb and \oiii lines.  \citeauthor{cid-fernandes2010} show that requiring reasonable \sn in all four of \hb, \oiii, \ha and \nii results in a 33 per decrease in sample size overall, and more than a factor of 2 decrease in the number of AGN+LINER systems relative to classification based on \ha and \nii alone.  For our sample, requiring detection of only \ha and \nii nearly doubles the number of classifiable emission-line systems.  In the bottom panel of Figure \ref{fig:figure11}, we plot EW(\ha) vs. \nii/\ha for galaxies in our full sample, where once again quenched galaxies are shown as filled contours.  Lines represent the so-called `optimal translation' of the classifications shown in the top panel of Figure \ref{fig:figure11} into the WHAN parameter space.  The observed surfeit of LINER-like systems suggested by the BPT classification is confirmed by the WHAN classification scheme, where the majority of systems are found to have relatively low EW(\ha).  However, as cautioned above, our ability to properly classify quenched Seyfert galaxies is hindered by estimates for the star-formation rates in these galaxies, which may be systematically overestimated by adopting \dfk as a proxy \citep[e.g.][]{salim2007}; our data therefore represents a lower limit on the true Seyfert fraction in such systems.

\section{Discussion}
\label{discussion}

We have used a combination of Galaxy Zoo visual morphologies and quantitative structural parameters to study the properties of quenched galaxies in the SDSS.  Our main result is that these quenched systems appear to host significant bulge components despite their young ages.

\subsection{The relationship between quenching and morphology}
\label{discussion_morph}

The buildup of the present-day red sequence can be understood as a combination of quenching processes and merging \citep[e.g.][]{faber2007}.  In this simplified picture, galaxies' recent merger history dictates their morphology---major mergers produce elliptical galaxies and minor mergers build spheroids without destroying stellar disks---while star formation is governed by a combination of feedback and environmental mechanisms. 

In this context, quenched galaxies provide leverage in disentangling the relative interplay between processes that govern galaxies' supply of cold gas and those that can alter their structure.  As discussed in Section \ref{vmorph}, Figures \ref{fig:figure9} and \ref{fig:figure10} show that quenched galaxies are both visually and structurally similar to other quiescent galaxies, despite their apparent recent arrival into the passive population.  Post-starburst galaxies, which represent an extreme tail of the quenched galaxy spectrum, paint a similar picture: \citet{quintero2004} show that post-starburst galaxies are more centrally concentrated (have higher S\'ersic indices) than star-forming galaxies, and \citet{balogh2005} show that while post-starburst galaxies host prominent bulges, their apparent progenitors---Balmer-strong emission galaxies, so-called e(a) systems \citep[e.g.][]{dressler1999}---are still largely disk dominated.  More generally, \citet{simard2009} showed that, at least in clusters, there is a correlation between the fraction of galaxies with \oii emission (i.e. star formation) and the fraction of galaxies with late-type morphologies.

It remains unclear, however, if the observed relationship between morphology and star-formation is causal---i.e. that a stellar bulge is \emph{required} for quenching to occur---or if the buildup of a stellar bulge is the result of physical processes that also induce quiescence.  Given the young ages of galaxies considered here, our results suggest that bulge formation \emph{must} precede the shutdown of star formation, i.e. that morphology is a critical factor in determining quiescence.  \citet{bell2012} have shown that there is a relationship between star formation and structure since at least $z\sim2$, where the buildup of a passive population is mirrored by the buildup of galaxies with high S\'ersic indices.  Similarly, \citet{wuyts2011} showed that at any given epoch there is a well defined relationship between galaxies' structure and their position on the star formation rate--stellar mass plane.  However, there is significant scatter between quiescence and any number of galaxy properties, suggesting that the mapping between structure and star formation is, at best, imperfect.  Indeed, there are notable exceptions to a paradigm where galaxies' star-formation activity is directly tied to their structure; passive spirals in particular pose a problem to any theory of galaxy evolution that limits the formation of passive galaxies to violent events, e.g. major mergers, and there is some evidence suggesting that quiescent disks may be quite common, particularly at $z \gtrsim 0.8$ \citep{bundy2010}.  However, both \citet{masters2010} and \citet{bundy2010} show that even passive disk galaxies appear to host more prominent bulges than their star forming counterparts, consistent with the results presented here.

\subsection{Where do post-starburst galaxies fit in?}
\label{psb_comp}

As mentioned previously, our selection of quenched systems builds upon techniques primarily used to identify post-starburst galaxies; it is therefore worthwhile to revisit the comparison between these two populations in light of the results discussed in Section \ref{results}.  By construction, the stringent criteria adopted for selecting post-starburst galaxies---e.g. EW(\hda) $\gtrsim$ 4--5\AA~and no \oii or \ha  emission---identify galaxies with extreme star-formation histories.  Of the 12,105 quenched galaxies in our sample, only 721 have \hda equivalent widths $\geq 4$\AA, or $\sim$0.1 per cent of the total SDSS sample, in relatively good agreement with other studies of post-starburst galaxies \citep{goto2003,blake2004,goto2005,wong2012}.

Several studies have highlighted the significant burst-mass fractions required to reproduce the absorption-line properties of local post-starburst galaxies. Based on fits to their $UV$+optical SEDs, \citet{kaviraj2007} find that 20--50 per cent of post-starburst galaxies' total stellar mass may be formed in recent bursts, while \citet{norton2001} and \citet{swinbank2012} derive more modest mass fractions of order 5 to 20 per cent.  By comparison, the inferred A-star mass fractions from our own fits are substantially lower, of order a few per cent, for the the majority of quenched galaxies.  In addition, we find a minority of galaxies with \fa greater than 10 per cent, and virtually none with \fa greater than 20 per cent.  It could be, then, that post-starburst galaxies exclusively trace very efficient star-formation activity---as may be the case in gas-rich major mergers \citep[e.g.][]{cox2006,saintonge2011a}---while the broader population of quenched galaxies host comparatively low-levels of star-formation prior to becoming quiescent.  In fact, merger-driven formation is also suggested by the high incidence of disturbed morphologies among post-starburst galaxies \citep{yang2004,blake2004,pracy2009}; the lack of a clear distinction between the local environments of post-starburst and `normal' galaxies lends further support to merging over more specific environmental process, e.g. ram-pressure stripping \citep{zabludoff1996,blake2004,hogg2006,wong2012}.

Yet, despite any differences between quenched and post-starburst galaxies in terms of their formation processes, the end products appear remarkably similar.  As discussed in Section \ref{discussion_morph}, one of the hallmarks of quenching appears to be the presence of a stellar bulge, and this holds equally true for post-starburst galaxies as it does for the more general quenched population \citep{quintero2004,balogh2005,goto2005,hogg2006,pracy2009}.  However, among quenched galaxies there is a clear correlation between \hda absorption-line width and central stellar mass surface density, pointing to increasingly efficient gas dissipation in the most extreme Balmer-line absorbers.  It may be, then, that true distinction between H$\delta$-strong post-starburst galaxies and the remainder of the quenched galaxy population lies in the \emph{efficiency} of bulge formation, rather than a particular formation mechanism.

\subsection{What quenches star formation?}
\label{discussion_quenching}

Ultimately, we would like to identify the process(es) that drive galaxies from star-forming to passive.  Given the relatively short lifetime of the quenching phase, ram-pressure stripping is an obvious culprit to account for the rapid decline of star-formation activity in quenched galaxies; however, it has been shown that simple gas stripping prescriptions dramatically overestimate the fraction of passive satellite galaxies in semi-analytic models \citep[e.g.][]{weinmann2006a,kimm2009}.  Furthermore, such simple models are inconsistent with the more gradual decline in star-formation activity inferred from statistical analyses of galaxy stellar populations \citep[e.g.][]{kauffmann2004}.  \citet{balogh2009} show that adopting a slower ``strangulation'' model for star formation as galaxies fall into dark matter haloes reduces some of the tension between observed and predicted colour distributions, but results in an over-production of green galaxies.  Recently, \citet{wetzel2012} have argued that the invariance of central and satellite SSFR distributions points towards a hybrid quenching model, where satellite galaxies evolve as normal for several Gyrs following accretion onto a massive dark matter halo and are ultimately quenched as a result of mergers, tidal interactions or harassment.  In fact, our data are consistent with any number of environmental mechanisms that may limit galaxies' gas supply without immediately truncating central star-formation activity; however, differentiating between such mechanisms requires a more detailed study of galaxies' environments that is beyond the scope of the present work.  Nevertheless, while the extent to which environment plays an important role in the evolution of ``average'' galaxies remains uncertain \citep[e.g.][]{peng2010}, the scarcity of pure disks in our sample (Figure \ref{fig:figure10}) suggests that even if \emph{all} passive galaxies ultimately form as the result of gas stripping processes in high-density environments, additional mechanisms above and beyond simple, wholesale gas removal are required to account for the correlation between bulge formation and quenching \citep[see also][]{bundy2010,bell2012}.  

Alternatively, galaxies' star formation may be regulated internally by energetic feedback from a central black hole \citep[e.g.][]{croton2006,hopkins2008,hopkins2008a}.  As discussed in Sections \ref{quenched_select} and \ref{agn}, our dependence on optical emission-line data to identify star-forming galaxies limits the extent to which we can study the association between optical AGN and quenching; however, it is still interesting to consider the relationship between quenched galaxies and other sub-samples with the understanding that the AGN fraction in quenched galaxies is a lower limit.  

We find that the fraction of quenched galaxies classified as AGN (i.e. having $S/N \geq 3$ in \hb, \oiii, \ha and \nii as well as satisfying the \citetalias{kauffmann2003} classification line) is comparable to that of other ``old'' galaxies\footnote{Here, the classification of galaxies as young or old is based on the distribution of \fa, i.e. Figure \ref{fig:figure2}}: 8 per cent of quenched galaxies are classified as AGN, compared to 17 and 7 per cent of young and old galaxies, respectively.  As discussed in Section \ref{agn}, the majority of AGN in quenched galaxies appear LINER-like (70--80 per cent) and have $L_\mathrm{[O\,{\scriptscriptstyle{III}}]} \leq 10^7~L_\odot$; in contrast, AGN in young galaxies are more than twice as likely to have high [\ion{O}{iii}] luminosities ($>$$10^7~L_\odot$) than either quenched or old systems. These results are in good agreement with previous studies showing that AGN are more frequently associated with recent or ongoing star formation \citep[e.g.][]{kauffmann2003,schawinski2007,best2012}.  We obtain a similar picture if we consider the frequency of high- and low-excitation radio-loud AGN using the recent catalogue of \citet{best2012}, which suggests that the paucity of AGN in quenched galaxies may be genuine \citep[see also][]{shin2011}.  It is interesting that we find no evidence for increase in the AGN activity of quenched galaxies, and hints that nuclear feedback may not be directly responsible for their transition from star forming to passive, however given the limitations of our sample we cannot make a more conclusive statement.  

Finally, \citet{martig2009} show that the addition of a dynamically-hot stellar component can suppress fragmentation in gas disks, so-called ``morphological quenching''.  In this scenario, the addition of a massive bulge results in a deepening of the central potential which, in turn, drives an increase in the epicyclic frequency of the disk; morphologically-quenched galaxies are therefore quenched \emph{despite} hosting significant gas disks.  Morphological quenching offers a mechanism to shutdown star formation in the absence of environmental and feedback effects, and can account for both the significant population of disks in our sample and the apparent lack of nuclear activity.  Such a mechanism may also account for the significant population of local early-type galaxies observed to host molecular gas \citep[22 per cent;][]{young2011} but little ongoing star formation.  If morphological quenching does contribute significantly to the suppression of star-formation in average galaxies, it remains important to understand why bulges appear to be a necessary but insufficient component of the quenching process (e.g. \citealp{bell2012}; see also discussion in \citealp{bundy2010}).

\section{Conclusions}

We have used the SDSS to identify the rare population of galaxies in transition between star forming and passive, and have used their photometric properties to study the mechanisms that govern star-formation quenching.  Our results can be summarised as follows:\\

\hangindent=0.3cm
\hangafter=0
\noindent i) Quenched galaxies are able to account for up to 60--70 per cent of the present day green-valley population after correcting for selection effects, suggesting that the majority of green galaxies will transition from star-forming to passive on $\sim$Gyr timescales.  However, the two samples select fundamentally different galaxy populations: while we require that quenched galaxies are passive, up to 80 per cent of ``green'' galaxies host significant emission.  \\

\hangindent=0.3cm
\hangafter=0
\noindent ii) Both visually and quantitatively, quenched galaxies appear to host significant bulge components despite their young ages.  It appears that the transition from star forming to passive is accompanied by significant morphological transformation, but that this transformation takes place in advance of or coincident with the shutdown of star formation.\\

\hangindent=0.3cm
\hangafter=0
\noindent iii) A comparison between quenched galaxies and classical post-starbursts shows that both populations conform to a scenario where bulge growth is an integral part of the quenching process.  Based on the comparison of their central stellar surface densities we suggest that the defining characteristic of post-starburst galaxies is the \emph{efficiency} of gas dissipation in the formation of their bulges, rather than a particular formation mechanism.\\

\hangindent=0.3cm
\hangafter=0
\noindent iv) Limitations of our sample selection notwithstanding, we find no evidence for an increase of the optical AGN fraction in quenched galaxies relative to other galaxies in the SDSS.  These results suggest that, if AGN \emph{do} play a decisive role in quenching star formation, then nuclear activity is relatively short lived compared to the observable lifetime of quenched galaxies.\\

Overall, our results show that galaxy quenching is closely tied to morphology, either through processes directly related to the presence of a bulge component \citep[e.g.][]{martig2009}, or indirectly through associated correlations with, e.g., black hole mass \citep[e.g.][]{hopkins2008}.  With our current data we are unable to identify the catalyst for bulge growth---which may be triggered internally via secular processes, or externally as the result of tidal interactions---however our results suggest that this process is a critical component governing the buildup of passive stellar mass with time.  Future multi-wavelength observations should help to distinguish between the multiple evolutionary pathways currently available to galaxies in theoretical models.

\section*{Acknowledgements}

SLE and DRP gratefully acknowledge the receipt of NSERC Discovery Grants which funded this research.  We also thank the anonymous referee, whose comments helped to improve this work.

Funding for the SDSS and SDSS-II has been provided by the Alfred P. Sloan Foundation, the Participating Institutions, the National Science Foundation, the U.S. Department of Energy, the National Aeronautics and Space Administration, the Japanese Monbukagakusho, the Max Planck Society, and the Higher Education Funding Council for England. The SDSS Web Site is http://www.sdss.org/.

The SDSS is managed by the Astrophysical Research Consortium for the Participating Institutions. The Participating Institutions are the American Museum of Natural History, Astrophysical Institute Potsdam, University of Basel, University of Cambridge, Case Western Reserve University, University of Chicago, Drexel University, Fermilab, the Institute for Advanced Study, the Japan Participation Group, Johns Hopkins University, the Joint Institute for Nuclear Astrophysics, the Kavli Institute for Particle Astrophysics and Cosmology, the Korean Scientist Group, the Chinese Academy of Sciences (LAMOST), Los Alamos National Laboratory, the Max-Planck-Institute for Astronomy (MPIA), the Max-Planck-Institute for Astrophysics (MPA), New Mexico State University, Ohio State University, University of Pittsburgh, University of Portsmouth, Princeton University, the United States Naval Observatory, and the University of Washington. 

\bibliographystyle{mn2e}
\bibliography{biblio}

\begin{thebibliography}{133}
\expandafter\ifx\csname natexlab\endcsname\relax\def\natexlab#1{#1}\fi

\bibitem[{{Abazajian} {et~al}\mbox{.}(2009){Abazajian}, {Adelman-McCarthy},
  {Ag{\"u}eros}, {Allam}, {Allende Prieto}, {An}, {Anderson}, {Anderson},
  {Annis}, {Bahcall}, {Bailer-Jones}, {Barentine}, {Bassett}, {Becker},
  {Beers}, {Bell}, {Belokurov}, {Berlind}, {Berman}, {Bernardi}, {Bickerton},
  {Bizyaev}, {Blakeslee}, {Blanton}, {Bochanski}, {Boroski}, {Brewington},
  {Brinchmann}, {Brinkmann}, {Brunner}, {Budav{\'a}ri}, {Carey}, {Carliles},
  {Carr}, {Castander}, {Cinabro}, {Connolly}, {Csabai}, {Cunha}, {Czarapata},
  {Davenport}, {de Haas}, {Dilday}, {Doi}, {Eisenstein}, {Evans}, {Evans},
  {Fan}, {Friedman}, {Frieman}, {Fukugita}, {G{\"a}nsicke}, {Gates},
  {Gillespie}, {Gilmore}, {Gonzalez}, {Gonzalez}, {Grebel}, {Gunn},
  {Gy{\"o}ry}, {Hall}, {Harding}, {Harris}, {Harvanek}, {Hawley}, {Hayes},
  {Heckman}, {Hendry}, {Hennessy}, {Hindsley}, {Hoblitt}, {Hogan}, {Hogg},
  {Holtzman}, {Hyde}, {Ichikawa}, {Ichikawa}, {Im}, {Ivezi{\'c}}, {Jester},
  {Jiang}, {Johnson}, {Jorgensen}, {Juri{\'c}}, {Kent}, {Kessler}, {Kleinman},
  {Knapp}, {Konishi}, {Kron}, {Krzesinski}, {Kuropatkin}, {Lampeitl},
  {Lebedeva}, {Lee}, {Lee}, {Leger}, {L{\'e}pine}, {Li}, {Lima}, {Lin}, {Long},
  {Loomis}, {Loveday}, {Lupton}, {Magnier}, {Malanushenko}, {Malanushenko},
  {Mandelbaum}, {Margon}, {Marriner}, {Mart{\'{\i}}nez-Delgado}, {Matsubara},
  {McGehee}, {McKay}, {Meiksin}, {Morrison}, {Mullally}, {Munn}, {Murphy},
  {Nash}, {Nebot}, {Neilsen}, {Newberg}, {Newman}, {Nichol}, {Nicinski},
  {Nieto-Santisteban}, {Nitta}, {Okamura}, {Oravetz}, {Ostriker}, {Owen},
  {Padmanabhan}, {Pan}, {Park}, {Pauls}, {Peoples}, {Percival}, {Pier}, {Pope},
  {Pourbaix}, {Price}, {Purger}, {Quinn}, {Raddick}, {Fiorentin}, {Richards},
  {Richmond}, {Riess}, {Rix}, {Rockosi}, {Sako}, {Schlegel}, {Schneider},
  {Scholz}, {Schreiber}, {Schwope}, {Seljak}, {Sesar}, {Sheldon}, {Shimasaku},
  {Sibley}, {Simmons}, {Sivarani}, {Smith}, {Smith}, {Smol{\v c}i{\'c}},
  {Snedden}, {Stebbins}, {Steinmetz}, {Stoughton}, {Strauss}, {Subba Rao},
  {Suto}, {Szalay}, {Szapudi}, {Szkody}, {Tanaka}, {Tegmark}, {Teodoro},
  {Thakar}, {Tremonti}, {Tucker}, {Uomoto}, {Vanden Berk}, {Vandenberg},
  {Vidrih}, {Vogeley}, {Voges}, {Vogt}, {Wadadekar}, {Watters}, {Weinberg},
  {West}, {White}, {Wilhite}, {Wonders}, {Yanny}, {Yocum}, {York}, {Zehavi},
  {Zibetti}, \& {Zucker}}]{abazajian2009}
{Abazajian} K.~N. {et~al.}, 2009, \apjs, 182, 543

\bibitem[{{Baldry} {et~al}\mbox{.}(2004){Baldry}, {Glazebrook}, {Brinkmann},
  {Ivezi{\'c}}, {Lupton}, {Nichol}, \& {Szalay}}]{baldry2004}
{Baldry} I.~K., {Glazebrook} K., {Brinkmann} J., {Ivezi{\'c}} {\v Z}., {Lupton}
  R.~H., {Nichol} R.~C., {Szalay} A.~S., 2004, \apj, 600, 681

\bibitem[{{Baldwin}, {Phillips} \& {Terlevich}(1981){Baldwin}, {Phillips}, \&
  {Terlevich}}]{baldwin1981}
{Baldwin} J.~A., {Phillips} M.~M., {Terlevich} R., 1981, \pasp, 93, 5

\bibitem[{{Balogh} {et~al}\mbox{.}(2004){Balogh}, {Baldry}, {Nichol}, {Miller},
  {Bower}, \& {Glazebrook}}]{balogh2004}
{Balogh} M.~L., {Baldry} I.~K., {Nichol} R., {Miller} C., {Bower} R.,
  {Glazebrook} K., 2004, \apjl, 615, L101

\bibitem[{{Balogh} {et~al}\mbox{.}(2009){Balogh}, {McGee}, {Wilman}, {Bower},
  {Hau}, {Morris}, {Mulchaey}, {Oemler}, {Parker}, \& {Gwyn}}]{balogh2009}
{Balogh} M.~L. {et~al.}, 2009, \mnras, 398, 754

\bibitem[{{Balogh} {et~al}\mbox{.}(2005){Balogh}, {Miller}, {Nichol},
  {Zabludoff}, \& {Goto}}]{balogh2005}
{Balogh} M.~L., {Miller} C., {Nichol} R., {Zabludoff} A., {Goto} T., 2005,
  \mnras, 360, 587

\bibitem[{{Balogh} {et~al}\mbox{.}(1999){Balogh}, {Morris}, {Yee}, {Carlberg},
  \& {Ellingson}}]{balogh1999}
{Balogh} M.~L., {Morris} S.~L., {Yee} H.~K.~C., {Carlberg} R.~G., {Ellingson}
  E., 1999, \apj, 527, 54

\bibitem[{{Bamford} {et~al}\mbox{.}(2009){Bamford}, {Nichol}, {Baldry}, {Land},
  {Lintott}, {Schawinski}, {Slosar}, {Szalay}, {Thomas}, {Torki}, {Andreescu},
  {Edmondson}, {Miller}, {Murray}, {Raddick}, \& {Vandenberg}}]{bamford2009}
{Bamford} S.~P. {et~al.}, 2009, \mnras, 393, 1324

\bibitem[{{Bell} {et~al}\mbox{.}(2012){Bell}, {van der Wel}, {Papovich},
  {Kocevski}, {Lotz}, {McIntosh}, {Kartaltepe}, {Faber}, {Ferguson},
  {Koekemoer}, {Grogin}, {Wuyts}, {Cheung}, {Conselice}, {Dekel}, {Dunlop},
  {Giavalisco}, {Herrington}, {Koo}, {McGrath}, {de Mello}, {Rix}, {Robaina},
  \& {Williams}}]{bell2012}
{Bell} E.~F. {et~al.}, 2012, \apj, 753, 167

\bibitem[{{Bell} {et~al}\mbox{.}(2004){Bell}, {Wolf}, {Meisenheimer}, {Rix},
  {Borch}, {Dye}, {Kleinheinrich}, {Wisotzki}, \& {McIntosh}}]{bell2004}
---, 2004, \apj, 608, 752

\bibitem[{{Bertin} \& {Arnouts}(1996)}]{bertin1996}
{Bertin} E., {Arnouts} S., 1996, \aaps, 117, 393

\bibitem[{{Best} \& {Heckman}(2012)}]{best2012}
{Best} P.~N., {Heckman} T.~M., 2012, \mnras, 421, 1569

\bibitem[{{Blake} {et~al}\mbox{.}(2004){Blake}, {Pracy}, {Couch}, {Bekki},
  {Lewis}, {Glazebrook}, {Baldry}, {Baugh}, {Bland-Hawthorn}, {Bridges},
  {Cannon}, {Cole}, {Colless}, {Collins}, {Dalton}, {De Propris}, {Driver},
  {Efstathiou}, {Ellis}, {Frenk}, {Jackson}, {Lahav}, {Lumsden}, {Maddox},
  {Madgwick}, {Norberg}, {Peacock}, {Peterson}, {Sutherland}, \&
  {Taylor}}]{blake2004}
{Blake} C. {et~al.}, 2004, \mnras, 355, 713

\bibitem[{{Blanton} \& {Roweis}(2007)}]{blanton2007a}
{Blanton} M.~R., {Roweis} S., 2007, \aj, 133, 734

\bibitem[{{Bower} {et~al}\mbox{.}(2006){Bower}, {Benson}, {Malbon}, {Helly},
  {Frenk}, {Baugh}, {Cole}, \& {Lacey}}]{bower2006}
{Bower} R.~G., {Benson} A.~J., {Malbon} R., {Helly} J.~C., {Frenk} C.~S.,
  {Baugh} C.~M., {Cole} S., {Lacey} C.~G., 2006, \mnras, 370, 645

\bibitem[{{Brammer} {et~al}\mbox{.}(2009){Brammer}, {Whitaker}, {van Dokkum},
  {Marchesini}, {Labb{\'e}}, {Franx}, {Kriek}, {Quadri}, {Illingworth}, {Lee},
  {Muzzin}, \& {Rudnick}}]{brammer2009}
{Brammer} G.~B. {et~al.}, 2009, \apjl, 706, L173

\bibitem[{{Brinchmann} {et~al}\mbox{.}(2004){Brinchmann}, {Charlot}, {White},
  {Tremonti}, {Kauffmann}, {Heckman}, \& {Brinkmann}}]{brinchmann2004}
{Brinchmann} J., {Charlot} S., {White} S.~D.~M., {Tremonti} C., {Kauffmann} G.,
  {Heckman} T., {Brinkmann} J., 2004, \mnras, 351, 1151

\bibitem[{{Bruzual} \& {Charlot}(2003)}]{bruzual2003}
{Bruzual} G., {Charlot} S., 2003, \mnras, 344, 1000

\bibitem[{{Bundy} {et~al}\mbox{.}(2010){Bundy}, {Scarlata}, {Carollo}, {Ellis},
  {Drory}, {Hopkins}, {Salvato}, {Leauthaud}, {Koekemoer}, {Murray}, {Ilbert},
  {Oesch}, {Ma}, {Capak}, {Pozzetti}, \& {Scoville}}]{bundy2010}
{Bundy} K. {et~al.}, 2010, \apj, 719, 1969

\bibitem[{{Chabrier}(2003)}]{chabrier2003}
{Chabrier} G., 2003, \pasp, 115, 763

\bibitem[{{Charlot} \& {Longhetti}(2001)}]{charlot2001}
{Charlot} S., {Longhetti} M., 2001, \mnras, 323, 887

\bibitem[{{Cid Fernandes} {et~al}\mbox{.}(2011){Cid Fernandes},
  {Stasi{\'n}ska}, {Mateus}, \& {Vale Asari}}]{cid-fernandes2011}
{Cid Fernandes} R., {Stasi{\'n}ska} G., {Mateus} A., {Vale Asari} N., 2011,
  \mnras, 413, 1687

\bibitem[{{Cid Fernandes} {et~al}\mbox{.}(2010){Cid Fernandes},
  {Stasi{\'n}ska}, {Schlickmann}, {Mateus}, {Vale Asari}, {Schoenell}, \&
  {Sodr{\'e}}}]{cid-fernandes2010}
{Cid Fernandes} R., {Stasi{\'n}ska} G., {Schlickmann} M.~S., {Mateus} A., {Vale
  Asari} N., {Schoenell} W., {Sodr{\'e}} L., 2010, \mnras, 403, 1036

\bibitem[{{Cirasuolo} {et~al}\mbox{.}(2007){Cirasuolo}, {McLure}, {Dunlop},
  {Almaini}, {Foucaud}, {Smail}, {Sekiguchi}, {Simpson}, {Eales}, {Dye},
  {Watson}, {Page}, \& {Hirst}}]{cirasuolo2007}
{Cirasuolo} M. {et~al.}, 2007, \mnras, 380, 585

\bibitem[{{Cole} {et~al}\mbox{.}(1994){Cole}, {Aragon-Salamanca}, {Frenk},
  {Navarro}, \& {Zepf}}]{cole1994}
{Cole} S., {Aragon-Salamanca} A., {Frenk} C.~S., {Navarro} J.~F., {Zepf} S.~E.,
  1994, \mnras, 271, 781

\bibitem[{{Colless} {et~al}\mbox{.}(2001){Colless}, {Dalton}, {Maddox},
  {Sutherland}, {Norberg}, {Cole}, {Bland-Hawthorn}, {Bridges}, {Cannon},
  {Collins}, {Couch}, {Cross}, {Deeley}, {De Propris}, {Driver}, {Efstathiou},
  {Ellis}, {Frenk}, {Glazebrook}, {Jackson}, {Lahav}, {Lewis}, {Lumsden},
  {Madgwick}, {Peacock}, {Peterson}, {Price}, {Seaborne}, \&
  {Taylor}}]{colless2001}
{Colless} M. {et~al.}, 2001, \mnras, 328, 1039

\bibitem[{{Conroy}, {Gunn} \& {White}(2009){Conroy}, {Gunn}, \&
  {White}}]{conroy2009}
{Conroy} C., {Gunn} J.~E., {White} M., 2009, \apj, 699, 486

\bibitem[{{Cooper} {et~al}\mbox{.}(2008){Cooper}, {Newman}, {Weiner}, {Yan},
  {Willmer}, {Bundy}, {Coil}, {Conselice}, {Davis}, {Faber}, {Gerke},
  {Guhathakurta}, {Koo}, \& {Noeske}}]{cooper2008a}
{Cooper} M.~C. {et~al.}, 2008, \mnras, 383, 1058

\bibitem[{{Couch} \& {Sharples}(1987)}]{couch1987}
{Couch} W.~J., {Sharples} R.~M., 1987, \mnras, 229, 423

\bibitem[{{Cox} {et~al}\mbox{.}(2006){Cox}, {Jonsson}, {Primack}, \&
  {Somerville}}]{cox2006}
{Cox} T.~J., {Jonsson} P., {Primack} J.~R., {Somerville} R.~S., 2006, \mnras,
  373, 1013

\bibitem[{{Croton} {et~al}\mbox{.}(2006){Croton}, {Springel}, {White}, {De
  Lucia}, {Frenk}, {Gao}, {Jenkins}, {Kauffmann}, {Navarro}, \&
  {Yoshida}}]{croton2006}
{Croton} D.~J. {et~al.}, 2006, \mnras, 365, 11

\bibitem[{{Daddi} {et~al}\mbox{.}(2007){Daddi}, {Dickinson}, {Morrison},
  {Chary}, {Cimatti}, {Elbaz}, {Frayer}, {Renzini}, {Pope}, {Alexander},
  {Bauer}, {Giavalisco}, {Huynh}, {Kurk}, \& {Mignoli}}]{daddi2007}
{Daddi} E. {et~al.}, 2007, \apj, 670, 156

\bibitem[{{Darg} {et~al}\mbox{.}(2010{\natexlab{a}}){Darg}, {Kaviraj},
  {Lintott}, {Schawinski}, {Sarzi}, {Bamford}, {Silk}, {Andreescu}, {Murray},
  {Nichol}, {Raddick}, {Slosar}, {Szalay}, {Thomas}, \&
  {Vandenberg}}]{darg2010a}
{Darg} D.~W. {et~al.}, 2010{\natexlab{a}}, \mnras, 401, 1552

\bibitem[{{Darg} {et~al}\mbox{.}(2010{\natexlab{b}}){Darg}, {Kaviraj},
  {Lintott}, {Schawinski}, {Sarzi}, {Bamford}, {Silk}, {Proctor}, {Andreescu},
  {Murray}, {Nichol}, {Raddick}, {Slosar}, {Szalay}, {Thomas}, \&
  {Vandenberg}}]{darg2010}
---, 2010{\natexlab{b}}, \mnras, 401, 1043

\bibitem[{{Davis} {et~al}\mbox{.}(2003){Davis}, {Faber}, {Newman}, {Phillips},
  {Ellis}, {Steidel}, {Conselice}, {Coil}, {Finkbeiner}, {Koo}, {Guhathakurta},
  {Weiner}, {Schiavon}, {Willmer}, {Kaiser}, {Luppino}, {Wirth}, {Connolly},
  {Eisenhardt}, {Cooper}, \& {Gerke}}]{davis2003}
{Davis} M. {et~al.}, 2003, in Society of Photo-Optical Instrumentation
  Engineers (SPIE) Conference Series, Vol. 4834, Society of Photo-Optical
  Instrumentation Engineers (SPIE) Conference Series, {P.~Guhathakurta}, ed.,
  pp. 161--172

\bibitem[{{De Lucia} {et~al}\mbox{.}(2007){De Lucia}, {Poggianti},
  {Arag{\'o}n-Salamanca}, {White}, {Zaritsky}, {Clowe}, {Halliday}, {Jablonka},
  {von der Linden}, {Milvang-Jensen}, {Pell{\'o}}, {Rudnick}, {Saglia}, \&
  {Simard}}]{de-lucia2007a}
{De Lucia} G. {et~al.}, 2007, \mnras, 374, 809

\bibitem[{{Dekel} \& {Birnboim}(2006)}]{dekel2006}
{Dekel} A., {Birnboim} Y., 2006, \mnras, 368, 2

\bibitem[{{Dressler} {et~al}\mbox{.}(1999){Dressler}, {Smail}, {Poggianti},
  {Butcher}, {Couch}, {Ellis}, \& {Oemler}}]{dressler1999}
{Dressler} A., {Smail} I., {Poggianti} B.~M., {Butcher} H., {Couch} W.~J.,
  {Ellis} R.~S., {Oemler}, Jr. A., 1999, \apjs, 122, 51

\bibitem[{{Ellingson} {et~al}\mbox{.}(2001){Ellingson}, {Lin}, {Yee}, \&
  {Carlberg}}]{ellingson2001}
{Ellingson} E., {Lin} H., {Yee} H.~K.~C., {Carlberg} R.~G., 2001, \apj, 547,
  609

\bibitem[{{Ellison} {et~al}\mbox{.}(2010){Ellison}, {Patton}, {Simard},
  {McConnachie}, {Baldry}, \& {Mendel}}]{ellison2010}
{Ellison} S.~L., {Patton} D.~R., {Simard} L., {McConnachie} A.~W., {Baldry}
  I.~K., {Mendel} J.~T., 2010, \mnras, 407, 1514

\bibitem[{{Faber} {et~al}\mbox{.}(2007){Faber}, {Willmer}, {Wolf}, {Koo},
  {Weiner}, {Newman}, {Im}, {Coil}, {Conroy}, {Cooper}, {Davis}, {Finkbeiner},
  {Gerke}, {Gebhardt}, {Groth}, {Guhathakurta}, {Harker}, {Kaiser}, {Kassin},
  {Kleinheinrich}, {Konidaris}, {Kron}, {Lin}, {Luppino}, {Madgwick},
  {Meisenheimer}, {Noeske}, {Phillips}, {Sarajedini}, {Schiavon}, {Simard},
  {Szalay}, {Vogt}, \& {Yan}}]{faber2007}
{Faber} S.~M. {et~al.}, 2007, \apj, 665, 265

\bibitem[{{Fontanot} {et~al}\mbox{.}(2009){Fontanot}, {De Lucia}, {Monaco},
  {Somerville}, \& {Santini}}]{fontanot2009}
{Fontanot} F., {De Lucia} G., {Monaco} P., {Somerville} R.~S., {Santini} P.,
  2009, \mnras, 397, 1776

\bibitem[{{Gallazzi} {et~al}\mbox{.}(2005){Gallazzi}, {Charlot}, {Brinchmann},
  {White}, \& {Tremonti}}]{gallazzi2005}
{Gallazzi} A., {Charlot} S., {Brinchmann} J., {White} S.~D.~M., {Tremonti}
  C.~A., 2005, \mnras, 362, 41

\bibitem[{{Goto}(2005)}]{goto2005}
{Goto} T., 2005, \mnras, 357, 937

\bibitem[{{Goto} {et~al}\mbox{.}(2003){Goto}, {Yamauchi}, {Fujita}, {Okamura},
  {Sekiguchi}, {Smail}, {Bernardi}, \& {Gomez}}]{goto2003}
{Goto} T., {Yamauchi} C., {Fujita} Y., {Okamura} S., {Sekiguchi} M., {Smail}
  I., {Bernardi} M., {Gomez} P.~L., 2003, \mnras, 346, 601

\bibitem[{{Goudfrooij} {et~al}\mbox{.}(1994){Goudfrooij}, {Hansen},
  {Jorgensen}, {Norgaard-Nielsen}, {de Jong}, \& {van den
  Hoek}}]{goudfrooij1994}
{Goudfrooij} P., {Hansen} L., {Jorgensen} H.~E., {Norgaard-Nielsen} H.~U., {de
  Jong} T., {van den Hoek} L.~B., 1994, \aaps, 104, 179

\bibitem[{{Gunn} \& {Gott}(1972)}]{gunn1972}
{Gunn} J.~E., {Gott}, III J.~R., 1972, \apj, 176, 1

\bibitem[{{Guo} {et~al}\mbox{.}(2011){Guo}, {White}, {Boylan-Kolchin}, {De
  Lucia}, {Kauffmann}, {Lemson}, {Li}, {Springel}, \& {Weinmann}}]{guo2011}
{Guo} Q. {et~al.}, 2011, \mnras, 413, 101

\bibitem[{{Henriques} {et~al}\mbox{.}(2009){Henriques}, {Thomas}, {Oliver}, \&
  {Roseboom}}]{henriques2009}
{Henriques} B.~M.~B., {Thomas} P.~A., {Oliver} S., {Roseboom} I., 2009, \mnras,
  396, 535

\bibitem[{{Hogg} {et~al}\mbox{.}(2006){Hogg}, {Masjedi}, {Berlind}, {Blanton},
  {Quintero}, \& {Brinkmann}}]{hogg2006}
{Hogg} D.~W., {Masjedi} M., {Berlind} A.~A., {Blanton} M.~R., {Quintero} A.~D.,
  {Brinkmann} J., 2006, \apj, 650, 763

\bibitem[{{Hopkins} {et~al}\mbox{.}(2008{\natexlab{a}}){Hopkins}, {Cox},
  {Kere{\v s}}, \& {Hernquist}}]{hopkins2008a}
{Hopkins} P.~F., {Cox} T.~J., {Kere{\v s}} D., {Hernquist} L.,
  2008{\natexlab{a}}, \apjs, 175, 390

\bibitem[{{Hopkins} {et~al}\mbox{.}(2009){Hopkins}, {Cox}, {Younger}, \&
  {Hernquist}}]{hopkins2009}
{Hopkins} P.~F., {Cox} T.~J., {Younger} J.~D., {Hernquist} L., 2009, \apj, 691,
  1168

\bibitem[{{Hopkins} {et~al}\mbox{.}(2008{\natexlab{b}}){Hopkins}, {Hernquist},
  {Cox}, \& {Kere{\v s}}}]{hopkins2008}
{Hopkins} P.~F., {Hernquist} L., {Cox} T.~J., {Kere{\v s}} D.,
  2008{\natexlab{b}}, \apjs, 175, 356

\bibitem[{{Kauffmann} \& {Haehnelt}(2000)}]{kauffmann2000}
{Kauffmann} G., {Haehnelt} M., 2000, \mnras, 311, 576

\bibitem[{{Kauffmann} {et~al}\mbox{.}(2003{\natexlab{a}}){Kauffmann},
  {Heckman}, {Tremonti}, {Brinchmann}, {Charlot}, {White}, {Ridgway},
  {Brinkmann}, {Fukugita}, {Hall}, {Ivezi{\'c}}, {Richards}, \&
  {Schneider}}]{kauffmann2003}
{Kauffmann} G. {et~al.}, 2003{\natexlab{a}}, \mnras, 346, 1055

\bibitem[{{Kauffmann} {et~al}\mbox{.}(2003{\natexlab{b}}){Kauffmann},
  {Heckman}, {White}, {Charlot}, {Tremonti}, {Brinchmann}, {Bruzual}, {Peng},
  {Seibert}, {Bernardi}, {Blanton}, {Brinkmann}, {Castander}, {Cs{\'a}bai},
  {Fukugita}, {Ivezic}, {Munn}, {Nichol}, {Padmanabhan}, {Thakar}, {Weinberg},
  \& {York}}]{kauffmann2003b}
---, 2003{\natexlab{b}}, \mnras, 341, 33

\bibitem[{{Kauffmann}, {White} \& {Guiderdoni}(1993){Kauffmann}, {White}, \&
  {Guiderdoni}}]{kauffmann1993}
{Kauffmann} G., {White} S.~D.~M., {Guiderdoni} B., 1993, \mnras, 264, 201

\bibitem[{{Kauffmann} {et~al}\mbox{.}(2004){Kauffmann}, {White}, {Heckman},
  {M{\'e}nard}, {Brinchmann}, {Charlot}, {Tremonti}, \&
  {Brinkmann}}]{kauffmann2004}
{Kauffmann} G., {White} S.~D.~M., {Heckman} T.~M., {M{\'e}nard} B.,
  {Brinchmann} J., {Charlot} S., {Tremonti} C., {Brinkmann} J., 2004, \mnras,
  353, 713

\bibitem[{{Kaviraj} {et~al}\mbox{.}(2007){Kaviraj}, {Kirkby}, {Silk}, \&
  {Sarzi}}]{kaviraj2007}
{Kaviraj} S., {Kirkby} L.~A., {Silk} J., {Sarzi} M., 2007, \mnras, 382, 960

\bibitem[{{Kere{\v s}} {et~al}\mbox{.}(2005){Kere{\v s}}, {Katz}, {Weinberg},
  \& {Dav{\'e}}}]{keres2005}
{Kere{\v s}} D., {Katz} N., {Weinberg} D.~H., {Dav{\'e}} R., 2005, \mnras, 363,
  2

\bibitem[{{Kewley} {et~al}\mbox{.}(2001){Kewley}, {Dopita}, {Sutherland},
  {Heisler}, \& {Trevena}}]{kewley2001}
{Kewley} L.~J., {Dopita} M.~A., {Sutherland} R.~S., {Heisler} C.~A., {Trevena}
  J., 2001, \apj, 556, 121

\bibitem[{{Kewley} {et~al}\mbox{.}(2006){Kewley}, {Groves}, {Kauffmann}, \&
  {Heckman}}]{kewley2006}
{Kewley} L.~J., {Groves} B., {Kauffmann} G., {Heckman} T., 2006, \mnras, 372,
  961

\bibitem[{{Kimm} {et~al}\mbox{.}(2009){Kimm}, {Somerville}, {Yi}, {van den
  Bosch}, {Salim}, {Fontanot}, {Monaco}, {Mo}, {Pasquali}, {Rich}, \&
  {Yang}}]{kimm2009}
{Kimm} T. {et~al.}, 2009, \mnras, 394, 1131

\bibitem[{{Kormendy} {et~al}\mbox{.}(2010){Kormendy}, {Drory}, {Bender}, \&
  {Cornell}}]{kormendy2010}
{Kormendy} J., {Drory} N., {Bender} R., {Cornell} M.~E., 2010, \apj, 723, 54

\bibitem[{{Lackner} \& {Gunn}(2012)}]{lackner2012}
{Lackner} C.~N., {Gunn} J.~E., 2012, \mnras, 421, 2277

\bibitem[{{Lewis} {et~al}\mbox{.}(2002){Lewis}, {Balogh}, {De Propris},
  {Couch}, {Bower}, {Offer}, {Bland-Hawthorn}, {Baldry}, {Baugh}, {Bridges},
  {Cannon}, {Cole}, {Colless}, {Collins}, {Cross}, {Dalton}, {Driver},
  {Efstathiou}, {Ellis}, {Frenk}, {Glazebrook}, {Hawkins}, {Jackson}, {Lahav},
  {Lumsden}, {Maddox}, {Madgwick}, {Norberg}, {Peacock}, {Percival},
  {Peterson}, {Sutherland}, \& {Taylor}}]{lewis2002}
{Lewis} I. {et~al.}, 2002, \mnras, 334, 673

\bibitem[{{Lilly} {et~al}\mbox{.}(2007){Lilly}, {Le F{\`e}vre}, {Renzini},
  {Zamorani}, {Scodeggio}, {Contini}, {Carollo}, {Hasinger}, {Kneib}, {Iovino},
  {Le Brun}, {Maier}, {Mainieri}, {Mignoli}, {Silverman}, {Tasca},
  {Bolzonella}, {Bongiorno}, {Bottini}, {Capak}, {Caputi}, {Cimatti},
  {Cucciati}, {Daddi}, {Feldmann}, {Franzetti}, {Garilli}, {Guzzo}, {Ilbert},
  {Kampczyk}, {Kovac}, {Lamareille}, {Leauthaud}, {Borgne}, {McCracken},
  {Marinoni}, {Pello}, {Ricciardelli}, {Scarlata}, {Vergani}, {Sanders},
  {Schinnerer}, {Scoville}, {Taniguchi}, {Arnouts}, {Aussel}, {Bardelli},
  {Brusa}, {Cappi}, {Ciliegi}, {Finoguenov}, {Foucaud}, {Franceschini},
  {Halliday}, {Impey}, {Knobel}, {Koekemoer}, {Kurk}, {Maccagni}, {Maddox},
  {Marano}, {Marconi}, {Meneux}, {Mobasher}, {Moreau}, {Peacock}, {Porciani},
  {Pozzetti}, {Scaramella}, {Schiminovich}, {Shopbell}, {Smail}, {Thompson},
  {Tresse}, {Vettolani}, {Zanichelli}, \& {Zucca}}]{lilly2007}
{Lilly} S.~J. {et~al.}, 2007, \apjs, 172, 70

\bibitem[{{Lintott} {et~al}\mbox{.}(2011){Lintott}, {Schawinski}, {Bamford},
  {Slosar}, {Land}, {Thomas}, {Edmondson}, {Masters}, {Nichol}, {Raddick},
  {Szalay}, {Andreescu}, {Murray}, \& {Vandenberg}}]{lintott2011}
{Lintott} C. {et~al.}, 2011, \mnras, 410, 166

\bibitem[{{Lintott} {et~al}\mbox{.}(2008){Lintott}, {Schawinski}, {Slosar},
  {Land}, {Bamford}, {Thomas}, {Raddick}, {Nichol}, {Szalay}, {Andreescu},
  {Murray}, \& {Vandenberg}}]{lintott2008}
{Lintott} C.~J. {et~al.}, 2008, \mnras, 389, 1179

\bibitem[{{Martig} {et~al}\mbox{.}(2009){Martig}, {Bournaud}, {Teyssier}, \&
  {Dekel}}]{martig2009}
{Martig} M., {Bournaud} F., {Teyssier} R., {Dekel} A., 2009, \apj, 707, 250

\bibitem[{{Martin} {et~al}\mbox{.}(2007){Martin}, {Wyder}, {Schiminovich},
  {Barlow}, {Forster}, {Friedman}, {Morrissey}, {Neff}, {Seibert}, {Small},
  {Welsh}, {Bianchi}, {Donas}, {Heckman}, {Lee}, {Madore}, {Milliard}, {Rich},
  {Szalay}, \& {Yi}}]{martin2007}
{Martin} D.~C. {et~al.}, 2007, \apjs, 173, 342

\bibitem[{{Masters} {et~al}\mbox{.}(2010){Masters}, {Mosleh}, {Romer},
  {Nichol}, {Bamford}, {Schawinski}, {Lintott}, {Andreescu}, {Campbell},
  {Crowcroft}, {Doyle}, {Edmondson}, {Murray}, {Raddick}, {Slosar}, {Szalay},
  \& {Vandenberg}}]{masters2010}
{Masters} K.~L. {et~al.}, 2010, \mnras, 405, 783

\bibitem[{{McGee} {et~al}\mbox{.}(2008){McGee}, {Balogh}, {Henderson},
  {Wilman}, {Bower}, {Mulchaey}, \& {Oemler}}]{mcgee2008}
{McGee} S.~L., {Balogh} M.~L., {Henderson} R.~D.~E., {Wilman} D.~J., {Bower}
  R.~G., {Mulchaey} J.~S., {Oemler}, Jr. A., 2008, \mnras, 387, 1605

\bibitem[{{McGee} {et~al}\mbox{.}(2011){McGee}, {Balogh}, {Wilman}, {Bower},
  {Mulchaey}, {Parker}, \& {Oemler}}]{mcgee2011}
{McGee} S.~L., {Balogh} M.~L., {Wilman} D.~J., {Bower} R.~G., {Mulchaey} J.~S.,
  {Parker} L.~C., {Oemler} A., 2011, \mnras, 413, 996

\bibitem[{{McIntosh} {et~al}\mbox{.}(2008){McIntosh}, {Guo}, {Hertzberg},
  {Katz}, {Mo}, {van den Bosch}, \& {Yang}}]{mcintosh2008}
{McIntosh} D.~H., {Guo} Y., {Hertzberg} J., {Katz} N., {Mo} H.~J., {van den
  Bosch} F.~C., {Yang} X., 2008, \mnras, 388, 1537

\bibitem[{{Mendel} {et~al}\mbox{.}(2012){Mendel}, {Palmer}, {Simard},
  {Ellison}, \& {Patton}}]{mendel2012a}
{Mendel} J.~T., {Palmer} M., {Simard} L., {Ellison} S.~L., {Patton} D.~R.,
  2012, \apjs, submitted

\bibitem[{{Mendez} {et~al}\mbox{.}(2011){Mendez}, {Coil}, {Lotz}, {Salim},
  {Moustakas}, \& {Simard}}]{mendez2011}
{Mendez} A.~J., {Coil} A.~L., {Lotz} J., {Salim} S., {Moustakas} J., {Simard}
  L., 2011, \apj, 736, 110

\bibitem[{{Mihos} \& {Hernquist}(1994)}]{mihos1994}
{Mihos} J.~C., {Hernquist} L., 1994, \apjl, 425, L13

\bibitem[{{Muzzin} {et~al}\mbox{.}(2012){Muzzin}, {Wilson}, {Yee}, {Gilbank},
  {Hoekstra}, {Demarco}, {Balogh}, {van Dokkum}, {Franx}, {Ellingson}, {Hicks},
  {Nantais}, {Noble}, {Lacy}, {Lidman}, {Rettura}, {Surace}, \&
  {Webb}}]{muzzin2012}
{Muzzin} A. {et~al.}, 2012, \apj, 746, 188

\bibitem[{{Negroponte} \& {White}(1983)}]{negroponte1983}
{Negroponte} J., {White} S.~D.~M., 1983, \mnras, 205, 1009

\bibitem[{{Noeske} {et~al}\mbox{.}(2007){Noeske}, {Weiner}, {Faber},
  {Papovich}, {Koo}, {Somerville}, {Bundy}, {Conselice}, {Newman},
  {Schiminovich}, {Le Floc'h}, {Coil}, {Rieke}, {Lotz}, {Primack}, {Barmby},
  {Cooper}, {Davis}, {Ellis}, {Fazio}, {Guhathakurta}, {Huang}, {Kassin},
  {Martin}, {Phillips}, {Rich}, {Small}, {Willmer}, \& {Wilson}}]{noeske2007}
{Noeske} K.~G. {et~al.}, 2007, \apjl, 660, L43

\bibitem[{{Nolan}, {Raychaudhury} \& {Kab{\'a}n}(2007){Nolan}, {Raychaudhury},
  \& {Kab{\'a}n}}]{nolan2007}
{Nolan} L.~A., {Raychaudhury} S., {Kab{\'a}n} A., 2007, \mnras, 375, 381

\bibitem[{{Norton} {et~al}\mbox{.}(2001){Norton}, {Gebhardt}, {Zabludoff}, \&
  {Zaritsky}}]{norton2001}
{Norton} S.~A., {Gebhardt} K., {Zabludoff} A.~I., {Zaritsky} D., 2001, \apj,
  557, 150

\bibitem[{{Nulsen}(1982)}]{nulsen1982}
{Nulsen} P.~E.~J., 1982, \mnras, 198, 1007

\bibitem[{{Peletier} {et~al}\mbox{.}(1990){Peletier}, {Davies}, {Illingworth},
  {Davis}, \& {Cawson}}]{peletier1990}
{Peletier} R.~F., {Davies} R.~L., {Illingworth} G.~D., {Davis} L.~E., {Cawson}
  M., 1990, \aj, 100, 1091

\bibitem[{{Peng} {et~al}\mbox{.}(2010){Peng}, {Lilly}, {Kova{\v c}},
  {Bolzonella}, {Pozzetti}, {Renzini}, {Zamorani}, {Ilbert}, {Knobel},
  {Iovino}, {Maier}, {Cucciati}, {Tasca}, {Carollo}, {Silverman}, {Kampczyk},
  {de Ravel}, {Sanders}, {Scoville}, {Contini}, {Mainieri}, {Scodeggio},
  {Kneib}, {Le F{\`e}vre}, {Bardelli}, {Bongiorno}, {Caputi}, {Coppa}, {de la
  Torre}, {Franzetti}, {Garilli}, {Lamareille}, {Le Borgne}, {Le Brun},
  {Mignoli}, {Perez Montero}, {Pello}, {Ricciardelli}, {Tanaka}, {Tresse},
  {Vergani}, {Welikala}, {Zucca}, {Oesch}, {Abbas}, {Barnes}, {Bordoloi},
  {Bottini}, {Cappi}, {Cassata}, {Cimatti}, {Fumana}, {Hasinger}, {Koekemoer},
  {Leauthaud}, {Maccagni}, {Marinoni}, {McCracken}, {Memeo}, {Meneux}, {Nair},
  {Porciani}, {Presotto}, \& {Scaramella}}]{peng2010}
{Peng} Y.-j. {et~al.}, 2010, \apj, 721, 193

\bibitem[{{Pracy} {et~al}\mbox{.}(2009){Pracy}, {Kuntschner}, {Couch}, {Blake},
  {Bekki}, \& {Briggs}}]{pracy2009}
{Pracy} M.~B., {Kuntschner} H., {Couch} W.~J., {Blake} C., {Bekki} K., {Briggs}
  F., 2009, \mnras, 396, 1349

\bibitem[{{Quintero} {et~al}\mbox{.}(2004){Quintero}, {Hogg}, {Blanton},
  {Schlegel}, {Eisenstein}, {Gunn}, {Brinkmann}, {Fukugita}, {Glazebrook}, \&
  {Goto}}]{quintero2004}
{Quintero} A.~D. {et~al.}, 2004, \apj, 602, 190

\bibitem[{{Roche}, {Bernardi} \& {Hyde}(2010){Roche}, {Bernardi}, \&
  {Hyde}}]{roche2010}
{Roche} N., {Bernardi} M., {Hyde} J., 2010, \mnras, 407, 1231

\bibitem[{{Saintonge} {et~al}\mbox{.}(2011){Saintonge}, {Kauffmann}, {Wang},
  {Kramer}, {Tacconi}, {Buchbender}, {Catinella}, {Graci{\'a}-Carpio},
  {Cortese}, {Fabello}, {Fu}, {Genzel}, {Giovanelli}, {Guo}, {Haynes},
  {Heckman}, {Krumholz}, {Lemonias}, {Li}, {Moran}, {Rodriguez-Fernandez},
  {Schiminovich}, {Schuster}, \& {Sievers}}]{saintonge2011a}
{Saintonge} A. {et~al.}, 2011, \mnras, 415, 61

\bibitem[{{Salim} {et~al}\mbox{.}(2007){Salim}, {Rich}, {Charlot},
  {Brinchmann}, {Johnson}, {Schiminovich}, {Seibert}, {Mallery}, {Heckman},
  {Forster}, {Friedman}, {Martin}, {Morrissey}, {Neff}, {Small}, {Wyder},
  {Bianchi}, {Donas}, {Lee}, {Madore}, {Milliard}, {Szalay}, {Welsh}, \&
  {Yi}}]{salim2007}
{Salim} S. {et~al.}, 2007, \apjs, 173, 267

\bibitem[{{Sarzi} {et~al}\mbox{.}(2010){Sarzi}, {Shields}, {Schawinski},
  {Jeong}, {Shapiro}, {Bacon}, {Bureau}, {Cappellari}, {Davies}, {de Zeeuw},
  {Emsellem}, {Falc{\'o}n-Barroso}, {Krajnovi{\'c}}, {Kuntschner}, {McDermid},
  {Peletier}, {van den Bosch}, {van de Ven}, \& {Yi}}]{sarzi2010}
{Sarzi} M. {et~al.}, 2010, \mnras, 402, 2187

\bibitem[{{Schawinski} {et~al}\mbox{.}(2007){Schawinski}, {Thomas}, {Sarzi},
  {Maraston}, {Kaviraj}, {Joo}, {Yi}, \& {Silk}}]{schawinski2007}
{Schawinski} K., {Thomas} D., {Sarzi} M., {Maraston} C., {Kaviraj} S., {Joo}
  S.-J., {Yi} S.~K., {Silk} J., 2007, \mnras, 382, 1415

\bibitem[{{Schlegel}, {Finkbeiner} \& {Davis}(1998){Schlegel}, {Finkbeiner}, \&
  {Davis}}]{schlegel1998}
{Schlegel} D.~J., {Finkbeiner} D.~P., {Davis} M., 1998, \apj, 500, 525

\bibitem[{{Schmidt}(1968)}]{schmidt1968}
{Schmidt} M., 1968, \apj, 151, 393

\bibitem[{{Scudder} {et~al}\mbox{.}(2012){Scudder}, {Ellison}, {Torrey},
  {Patton}, \& {Mendel}}]{scudder2012a}
{Scudder} J.~M., {Ellison} S.~L., {Torrey} P., {Patton} D.~R., {Mendel} J.~T.,
  2012, \mnras, 426, 549

\bibitem[{{Sellwood} \& {Wilkinson}(1993)}]{sellwood1993}
{Sellwood} J.~A., {Wilkinson} A., 1993, Reports on Progress in Physics, 56, 173

\bibitem[{{Shin}, {Strauss} \& {Tojeiro}(2011){Shin}, {Strauss}, \&
  {Tojeiro}}]{shin2011}
{Shin} M.-S., {Strauss} M.~A., {Tojeiro} R., 2011, \mnras, 410, 1583

\bibitem[{{Simard} {et~al}\mbox{.}(2009){Simard}, {Clowe}, {Desai},
  {Dalcanton}, {von der Linden}, {Poggianti}, {White}, {Arag{\'o}n-Salamanca},
  {De Lucia}, {Halliday}, {Jablonka}, {Milvang-Jensen}, {Saglia}, {Pell{\'o}},
  {Rudnick}, \& {Zaritsky}}]{simard2009}
{Simard} L. {et~al.}, 2009, \aap, 508, 1141

\bibitem[{{Simard} {et~al}\mbox{.}(2011){Simard}, {Mendel}, {Patton},
  {Ellison}, \& {McConnachie}}]{simard2011}
{Simard} L., {Mendel} J.~T., {Patton} D.~R., {Ellison} S.~L., {McConnachie}
  A.~W., 2011, \apjs, 196, 11

\bibitem[{{Simard} {et~al}\mbox{.}(2002){Simard}, {Willmer}, {Vogt},
  {Sarajedini}, {Phillips}, {Weiner}, {Koo}, {Im}, {Illingworth}, \&
  {Faber}}]{simard2002}
{Simard} L. {et~al.}, 2002, \apjs, 142, 1

\bibitem[{{Skibba} {et~al}\mbox{.}(2009){Skibba}, {Bamford}, {Nichol},
  {Lintott}, {Andreescu}, {Edmondson}, {Murray}, {Raddick}, {Schawinski},
  {Slosar}, {Szalay}, {Thomas}, \& {Vandenberg}}]{skibba2009}
{Skibba} R.~A. {et~al.}, 2009, \mnras, 399, 966

\bibitem[{{Snyder} {et~al}\mbox{.}(2011){Snyder}, {Cox}, {Hayward},
  {Hernquist}, \& {Jonsson}}]{snyder2011}
{Snyder} G.~F., {Cox} T.~J., {Hayward} C.~C., {Hernquist} L., {Jonsson} P.,
  2011, \apj, 741, 77

\bibitem[{{Somerville} {et~al}\mbox{.}(2008){Somerville}, {Hopkins}, {Cox},
  {Robertson}, \& {Hernquist}}]{somerville2008}
{Somerville} R.~S., {Hopkins} P.~F., {Cox} T.~J., {Robertson} B.~E.,
  {Hernquist} L., 2008, \mnras, 391, 481

\bibitem[{{Somerville} \& {Primack}(1999)}]{somerville1999}
{Somerville} R.~S., {Primack} J.~R., 1999, \mnras, 310, 1087

\bibitem[{{Stasi{\'n}ska} {et~al}\mbox{.}(2006){Stasi{\'n}ska}, {Cid
  Fernandes}, {Mateus}, {Sodr{\'e}}, \& {Asari}}]{stasinska2006}
{Stasi{\'n}ska} G., {Cid Fernandes} R., {Mateus} A., {Sodr{\'e}} L., {Asari}
  N.~V., 2006, \mnras, 371, 972

\bibitem[{{Stasi{\'n}ska} {et~al}\mbox{.}(2008){Stasi{\'n}ska}, {Vale Asari},
  {Cid Fernandes}, {Gomes}, {Schlickmann}, {Mateus}, {Schoenell}, \&
  {Sodr{\'e}}}]{stasinska2008}
{Stasi{\'n}ska} G., {Vale Asari} N., {Cid Fernandes} R., {Gomes} J.~M.,
  {Schlickmann} M., {Mateus} A., {Schoenell} W., {Sodr{\'e}},
  Jr.~Seagal~Collaboration L., 2008, \mnras, 391, L29

\bibitem[{{Strateva} {et~al}\mbox{.}(2001){Strateva}, {Ivezi{\'c}}, {Knapp},
  {Narayanan}, {Strauss}, {Gunn}, {Lupton}, {Schlegel}, {Bahcall}, {Brinkmann},
  {Brunner}, {Budav{\'a}ri}, {Csabai}, {Castander}, {Doi}, {Fukugita}, {Gy{\H
  o}ry}, {Hamabe}, {Hennessy}, {Ichikawa}, {Kunszt}, {Lamb}, {McKay},
  {Okamura}, {Racusin}, {Sekiguchi}, {Schneider}, {Shimasaku}, \&
  {York}}]{strateva2001}
{Strateva} I. {et~al.}, 2001, \aj, 122, 1861

\bibitem[{{Swinbank} {et~al}\mbox{.}(2012){Swinbank}, {Balogh}, {Bower},
  {Zabludoff}, {Lucey}, {McGee}, {Miller}, \& {Nichol}}]{swinbank2012}
{Swinbank} A.~M., {Balogh} M.~L., {Bower} R.~G., {Zabludoff} A.~I., {Lucey}
  J.~R., {McGee} S.~L., {Miller} C.~J., {Nichol} R.~C., 2012, \mnras, 420, 672

\bibitem[{{Taniguchi}, {Shioya} \& {Murayama}(2000){Taniguchi}, {Shioya}, \&
  {Murayama}}]{taniguchi2000}
{Taniguchi} Y., {Shioya} Y., {Murayama} T., 2000, \aj, 120, 1265

\bibitem[{{Toomre} \& {Toomre}(1972)}]{toomre1972}
{Toomre} A., {Toomre} J., 1972, \apj, 178, 623

\bibitem[{{van de Voort} {et~al}\mbox{.}(2011){van de Voort}, {Schaye},
  {Booth}, \& {Dalla Vecchia}}]{van-de-voort2011}
{van de Voort} F., {Schaye} J., {Booth} C.~M., {Dalla Vecchia} C., 2011,
  \mnras, 415, 2782

\bibitem[{{van den Bosch} {et~al}\mbox{.}(2008){van den Bosch}, {Aquino},
  {Yang}, {Mo}, {Pasquali}, {McIntosh}, {Weinmann}, \&
  {Kang}}]{van-den-bosch2008}
{van den Bosch} F.~C., {Aquino} D., {Yang} X., {Mo} H.~J., {Pasquali} A.,
  {McIntosh} D.~H., {Weinmann} S.~M., {Kang} X., 2008, \mnras, 387, 79

\bibitem[{{von der Linden} {et~al}\mbox{.}(2010){von der Linden}, {Wild},
  {Kauffmann}, {White}, \& {Weinmann}}]{von-der-linden2010}
{von der Linden} A., {Wild} V., {Kauffmann} G., {White} S.~D.~M., {Weinmann}
  S., 2010, \mnras, 404, 1231

\bibitem[{{Weinmann} {et~al}\mbox{.}(2012){Weinmann}, {Pasquali},
  {Oppenheimer}, {Finlator}, {Mendel}, {Crain}, \& {Macci{\`o}}}]{weinmann2012}
{Weinmann} S.~M., {Pasquali} A., {Oppenheimer} B.~D., {Finlator} K., {Mendel}
  J.~T., {Crain} R.~A., {Macci{\`o}} A.~V., 2012, \mnras, 426, 2797

\bibitem[{{Weinmann} {et~al}\mbox{.}(2006{\natexlab{a}}){Weinmann}, {van den
  Bosch}, {Yang}, \& {Mo}}]{weinmann2006}
{Weinmann} S.~M., {van den Bosch} F.~C., {Yang} X., {Mo} H.~J.,
  2006{\natexlab{a}}, \mnras, 366, 2

\bibitem[{{Weinmann} {et~al}\mbox{.}(2006{\natexlab{b}}){Weinmann}, {van den
  Bosch}, {Yang}, {Mo}, {Croton}, \& {Moore}}]{weinmann2006a}
{Weinmann} S.~M., {van den Bosch} F.~C., {Yang} X., {Mo} H.~J., {Croton} D.~J.,
  {Moore} B., 2006{\natexlab{b}}, \mnras, 372, 1161

\bibitem[{{Wetzel}, {Tinker} \& {Conroy}(2012){Wetzel}, {Tinker}, \&
  {Conroy}}]{wetzel2012}
{Wetzel} A.~R., {Tinker} J.~L., {Conroy} C., 2012, \mnras, 424, 232

\bibitem[{{Wild} {et~al}\mbox{.}(2007){Wild}, {Kauffmann}, {Heckman},
  {Charlot}, {Lemson}, {Brinchmann}, {Reichard}, \& {Pasquali}}]{wild2007}
{Wild} V., {Kauffmann} G., {Heckman} T., {Charlot} S., {Lemson} G.,
  {Brinchmann} J., {Reichard} T., {Pasquali} A., 2007, \mnras, 381, 543

\bibitem[{{Wild} {et~al}\mbox{.}(2009){Wild}, {Walcher}, {Johansson}, {Tresse},
  {Charlot}, {Pollo}, {Le F{\`e}vre}, \& {de Ravel}}]{wild2009}
{Wild} V., {Walcher} C.~J., {Johansson} P.~H., {Tresse} L., {Charlot} S.,
  {Pollo} A., {Le F{\`e}vre} O., {de Ravel} L., 2009, \mnras, 395, 144

\bibitem[{{Williams} {et~al}\mbox{.}(2009){Williams}, {Quadri}, {Franx}, {van
  Dokkum}, \& {Labb{\'e}}}]{williams2009}
{Williams} R.~J., {Quadri} R.~F., {Franx} M., {van Dokkum} P., {Labb{\'e}} I.,
  2009, \apj, 691, 1879

\bibitem[{{Wong} {et~al}\mbox{.}(2012){Wong}, {Schawinski}, {Kaviraj},
  {Masters}, {Nichol}, {Lintott}, {Keel}, {Darg}, {Bamford}, {Andreescu},
  {Murray}, {Raddick}, {Szalay}, {Thomas}, \& {Vandenberg}}]{wong2012}
{Wong} O.~I. {et~al.}, 2012, \mnras, 420, 1684

\bibitem[{{Woo} {et~al}\mbox{.}(2012){Woo}, {Dekel}, {Faber}, {Noeske}, {Koo},
  {Gerke}, {Cooper}, {Salim}, {Dutton}, {Newman}, {Weiner}, {Bundy}, {Willmer},
  {Davis}, \& {Yan}}]{woo2012}
{Woo} J. {et~al.}, 2012, ArXiv e-prints

\bibitem[{{Worthey} \& {Ottaviani}(1997)}]{worthey1997}
{Worthey} G., {Ottaviani} D.~L., 1997, \apjs, 111, 377

\bibitem[{{Wuyts} {et~al}\mbox{.}(2011){Wuyts}, {F{\"o}rster Schreiber}, {van
  der Wel}, {Magnelli}, {Guo}, {Genzel}, {Lutz}, {Aussel}, {Barro}, {Berta},
  {Cava}, {Graci{\'a}-Carpio}, {Hathi}, {Huang}, {Kocevski}, {Koekemoer},
  {Lee}, {Le Floc'h}, {McGrath}, {Nordon}, {Popesso}, {Pozzi}, {Riguccini},
  {Rodighiero}, {Saintonge}, \& {Tacconi}}]{wuyts2011}
{Wuyts} S. {et~al.}, 2011, \apj, 742, 96

\bibitem[{{Yan} \& {Blanton}(2012)}]{yan2012}
{Yan} R., {Blanton} M.~R., 2012, \apj, 747, 61

\bibitem[{{Yan} {et~al}\mbox{.}(2009){Yan}, {Newman}, {Faber}, {Coil},
  {Cooper}, {Davis}, {Weiner}, {Gerke}, \& {Koo}}]{yan2009}
{Yan} R. {et~al.}, 2009, \mnras, 398, 735

\bibitem[{{Yan} {et~al}\mbox{.}(2006){Yan}, {Newman}, {Faber}, {Konidaris},
  {Koo}, \& {Davis}}]{yan2006}
{Yan} R., {Newman} J.~A., {Faber} S.~M., {Konidaris} N., {Koo} D., {Davis} M.,
  2006, \apj, 648, 281

\bibitem[{{Yang} {et~al}\mbox{.}(2004{\natexlab{a}}){Yang}, {Mo}, {Jing}, {van
  den Bosch}, \& {Chu}}]{yang2004}
{Yang} X., {Mo} H.~J., {Jing} Y.~P., {van den Bosch} F.~C., {Chu} Y.,
  2004{\natexlab{a}}, \mnras, 350, 1153

\bibitem[{{Yang} {et~al}\mbox{.}(2004{\natexlab{b}}){Yang}, {Zabludoff},
  {Zaritsky}, {Lauer}, \& {Mihos}}]{yang2004a}
{Yang} Y., {Zabludoff} A.~I., {Zaritsky} D., {Lauer} T.~R., {Mihos} J.~C.,
  2004{\natexlab{b}}, \apj, 607, 258

\bibitem[{{York} {et~al}\mbox{.}(2000){York}, {Adelman}, {Anderson},
  {Anderson}, {Annis}, {Bahcall}, {Bakken}, {Barkhouser}, {Bastian}, {Berman},
  {Boroski}, {Bracker}, {Briegel}, {Briggs}, {Brinkmann}, {Brunner}, {Burles},
  {Carey}, {Carr}, {Castander}, {Chen}, {Colestock}, {Connolly}, {Crocker},
  {Csabai}, {Czarapata}, {Davis}, {Doi}, {Dombeck}, {Eisenstein}, {Ellman},
  {Elms}, {Evans}, {Fan}, {Federwitz}, {Fiscelli}, {Friedman}, {Frieman},
  {Fukugita}, {Gillespie}, {Gunn}, {Gurbani}, {de Haas}, {Haldeman}, {Harris},
  {Hayes}, {Heckman}, {Hennessy}, {Hindsley}, {Holm}, {Holmgren}, {Huang},
  {Hull}, {Husby}, {Ichikawa}, {Ichikawa}, {Ivezi{\'c}}, {Kent}, {Kim},
  {Kinney}, {Klaene}, {Kleinman}, {Kleinman}, {Knapp}, {Korienek}, {Kron},
  {Kunszt}, {Lamb}, {Lee}, {Leger}, {Limmongkol}, {Lindenmeyer}, {Long},
  {Loomis}, {Loveday}, {Lucinio}, {Lupton}, {MacKinnon}, {Mannery}, {Mantsch},
  {Margon}, {McGehee}, {McKay}, {Meiksin}, {Merelli}, {Monet}, {Munn},
  {Narayanan}, {Nash}, {Neilsen}, {Neswold}, {Newberg}, {Nichol}, {Nicinski},
  {Nonino}, {Okada}, {Okamura}, {Ostriker}, {Owen}, {Pauls}, {Peoples},
  {Peterson}, {Petravick}, {Pier}, {Pope}, {Pordes}, {Prosapio},
  {Rechenmacher}, {Quinn}, {Richards}, {Richmond}, {Rivetta}, {Rockosi},
  {Ruthmansdorfer}, {Sandford}, {Schlegel}, {Schneider}, {Sekiguchi}, {Sergey},
  {Shimasaku}, {Siegmund}, {Smee}, {Smith}, {Snedden}, {Stone}, {Stoughton},
  {Strauss}, {Stubbs}, {SubbaRao}, {Szalay}, {Szapudi}, {Szokoly}, {Thakar},
  {Tremonti}, {Tucker}, {Uomoto}, {Vanden Berk}, {Vogeley}, {Waddell}, {Wang},
  {Watanabe}, {Weinberg}, {Yanny}, \& {Yasuda}}]{york2000}
{York} D.~G. {et~al.}, 2000, \aj, 120, 1579

\bibitem[{{Young} {et~al}\mbox{.}(2011){Young}, {Bureau}, {Davis}, {Combes},
  {McDermid}, {Alatalo}, {Blitz}, {Bois}, {Bournaud}, {Cappellari}, {Davies},
  {de Zeeuw}, {Emsellem}, {Khochfar}, {Krajnovi{\'c}}, {Kuntschner},
  {Lablanche}, {Morganti}, {Naab}, {Oosterloo}, {Sarzi}, {Scott}, {Serra}, \&
  {Weijmans}}]{young2011}
{Young} L.~M. {et~al.}, 2011, \mnras, 414, 940

\bibitem[{{Zabludoff} {et~al}\mbox{.}(1996){Zabludoff}, {Zaritsky}, {Lin},
  {Tucker}, {Hashimoto}, {Shectman}, {Oemler}, \& {Kirshner}}]{zabludoff1996}
{Zabludoff} A.~I., {Zaritsky} D., {Lin} H., {Tucker} D., {Hashimoto} Y.,
  {Shectman} S.~A., {Oemler} A., {Kirshner} R.~P., 1996, \apj, 466, 104

\end{thebibliography}

\end{document}